\documentclass[12pt]{article}
\usepackage{graphicx} % Required for inserting images

\usepackage{hepnames}
\usepackage[printonlyused,nohyperlinks]{acronym}
\usepackage{comment,fullpage}
\usepackage{rotating}

\usepackage{booktabs,colortbl,multirow,amssymb}
\usepackage{arydshln}
\usepackage{url}

\newcommand{\al}{\alpha}
\newcommand{\as}{\alpha_{\rm s}}
\newcommand{\af}{\alpha_{\rm f}}
\newcommand{\at}{\alpha_{\rm t}}

\newcommand{\crowcolorC}{\rowcolor[rgb]{0.9,0.9,0.9}}

\newcommand{\cellcolorC}{\cellcolor[rgb]{0.9,0.9,0.9}}
\newcommand{\cellcolorCurr}{\cellcolor[rgb]{0.8,0.8,0.8}}

\newcommand{\vtext}[1]{\begin{sideways}\small{#1}\end{sideways}}

\title{Theoretical and phenomenological aspects of the\\  Electroweak physics WG studies for the\\  2026 ESPPU Physics Briefing Book}
\author{J. de Blas$^1$, M.~Dunford$^2$ (Conveners),\\ E. Bagnaschi$^3$ (Scientific secretary), A. Freitas$^4$, P. P. Giardino$^5$, \\ C. Grefe$^6$, M. Selvaggi$^7$, A. Taliercio$^8$, F. Bartels$^7$}
\date{September 2026}

\begin{document}

\begin{flushright} IFT-UAM/CSIC-26-103 \end{flushright}

{
\let\newpage\relax
\maketitle
}

\maketitle

\begin{center}
$^1$ {\em Universidad de Granada, Spain}\\
$^2$ {\em Universit\"at Heidelberg, Germany}\\
$^3$ {\em INFN Laboratori Nazionali di Frascati, Italy}\\
$^4$ {\em University of Pittsburgh, US}\\
$^5$ {\em Universidad Aut\'onoma de Madrid, Spain} \\
$^6$ {\em Universit\"at Bonn, Germany}\\
$^7$ {\em CERN, Geneva, Switzerland}\\
$^8$ {\em Northwestern University, US}
\end{center}

\section{Introduction}

The 2025–2026 update of the European Strategy for Particle Physics placed renewed emphasis on quantitatively robust comparisons between future collider projects. Within this context, the \ac{PPG EW WG} carried out a comprehensive assessment of the projected sensitivities of different facilities to Higgs boson, electroweak and top-quark observables, with particular attention on different scenarios of theoretical uncertainties. These studies are summarized in the Physics Briefing Book~\cite{deBlas:2025PhysicsBriefingBook}.

The aim of this document is to (a) provide a more detailed account of the evaluation of theoretical uncertainties, which was undertaken within the \ac{PPG EW WG}, and (b)  document the collider-specific inputs used in the electroweak and \ac{SMEFT} fits and discuss any needed modifications in order to harmonize assumptions between projects. In presenting the detailed theoretical uncertainty treatment and the explicit collider inputs and assumptions, this work aims to improve the reproducibility and future extensibility of the results presented in the Physics Briefing Book.

\section{Introduction on theory uncertainties}

The interpretation of precision measurements in particle physics relies crucially on the accuracy of the underlying theoretical predictions. An essential component of this accuracy is the control and assessment of \emph{theory uncertainties}, an umbrella term that encompasses a variety of uncertainties associated with the theoretical and modeling framework used in a given analysis, as well as with its physical interpretation. These uncertainties reflect limitations of perturbative calculations, modeling assumptions, external inputs, and approximations inherent in our current theoretical descriptions.

By their very nature, theory uncertainties cannot always be defined or quantified in a strictly statistical manner. Their estimation therefore involves a certain degree of arbitrariness, which distinguishes them from experimental uncertainties. For this reason, a central goal of precision phenomenology is to ensure that theory uncertainties remain subdominant compared to experimental ones. Achieving this goal is particularly challenging in the context of future high-precision measurements and will require sustained and coordinated theoretical effort.

The purpose of this section is to review, update, and extend existing estimates of theory uncertainties, with a primary focus on observables relevant for $e^+e^-$ collider physics. We summarize the current state of theoretical control and discuss possible scenarios for future improvements. Rather than attempting to predict which of these scenarios is most likely to be realized, or to define a strict theoretical limit on achievable precision, our aim is to estimate the level of theoretical improvement required in order to match a given target experimental accuracy.

Theory uncertainties arise from several distinct sources. One of the most prominent contributions is due to missing higher-order uncertainties (MHOUs), originating from the truncation of perturbative expansions at finite order. Additional contributions stem from parton distribution functions (PDFs), whose uncertainties reflect not only the experimental precision of the data entering the fits, but also theoretical uncertainties in the fitted observables and modeling choices such as parameterizations and heavy-quark treatments. Further sources include non-perturbative (NP) effects, such as hadronization and low-scale dynamics, which are typically described using physics-inspired models and constrained by data, as well as parametric uncertainties associated with the determination of external input parameters (e.g.\ masses or couplings) that themselves require theoretical input for their extraction.

Due to their central role, several complementary strategies are commonly employed to estimate the impact of MHOUs. A first approach is based on power counting and the size of perturbative prefactors, for example,
\begin{align}
  \frac{g^2}{4\pi^2}\,n_f \quad \text{for electroweak loop corrections}, \qquad
  \frac{\alpha_s}{\pi}\,C \quad \text{for QCD loop corrections},
\label{eq:prefac}
\end{align}
where $n_f$ denotes the number of fermion species and $C$ the relevant Casimir factor. A second method relies on extrapolating the perturbative series, often assuming a geometric behavior of successive terms, for instance,
\begin{align}
  {\cal O}(\alpha^3)-{\cal O}(\tilde{\alpha}^3)
  \;\approx\;
  \frac{{\cal O}(\alpha^2)-{\cal O}(\tilde{\alpha}^2)}{{\cal O}(\alpha)}\,
  {\cal O}(\alpha^2), \label{eq:geom}
\end{align}
with $\tilde{\alpha}$ denoting an approximate calculation of the given order, e.g. in the large top-mass limit. In this formula it is assumed that all orders except ${\cal O}(\alpha^3)$ are known and the impact of the latter is being estimated. 

A widely used and complementary estimate is obtained by varying the renormalization scale in $\overline{\mathrm{MS}}$ quantities, thereby probing the sensitivity to missing higher orders. For $Z$-pole observables, a typical choice is
\[
  \mu = m_Z,\; \frac{m_Z}{2},\; 2 m_Z.
\]
Another possibility is the comparison between different renormalization schemes, such as on-shell versus $\overline{\mathrm{MS}}$, which provides complementary insight into the potential size of non-computed higher-order contributions.

More complex approaches have also been recently introduced, with the aim of better integrating the theory uncertainties in statistical fits. 
Among these new developments, we briefly mention: the possibility of using series acceleration~\cite{David:2013gaa}; the introduction of Bayesian modeling, as first proposed in in Ref.~\cite{Cacciari:2011ze}, and subsequently further developed in several studies~\cite{Forte:2013mda, Bagnaschi:2014wea, Bonvini:2020xeo, Duhr:2021mfd}; the introduction, and experimental profiling, of the so-called ``Theory Nuisance Parameters''~\cite{Tackmann:2024kci,Lim:2024nsk}.

Non-perturbative uncertainties are usually assessed using a combination of approaches, including variations of model parameters, comparisons between different phenomenological models and, increasingly, first-principles lattice QCD calculations with a systematic evaluation of their uncertainties. Together, these methods aim to provide a coherent and conservative estimate of the impact of non-perturbative effects on precision observables.

This framework provides the basis for the theory uncertainty estimates discussed in the remainder of this paper and for assessing the theoretical requirements needed to fully exploit the physics potential of future precision experiments.

\subsection{Theory uncertainties for \boldmath $e^+e^-$ colliders}

At $e^+e^-$ colliders, the \ac{EWPO}s and Higgs precision observables (HPOs) pose the highest demands on precision theory input. Although the current report provides estimates for a larger range of error sources than in previous works, it is nevertheless not comprehensive.\footnote{In particular, we do not evaluate theory errors in the determination of the beam energy and luminosity~\cite{Altmann:2025feg}.} Several 
scenarios are considered: 
\begin{itemize}
\item[i)] current status (i.e.~evaluation of uncertainties of existing calculations);
\item[ii)] a ``conservative'' scenario, assuming theory improvements which are likely to be achieved by building on and extending existing computational methods;
\item[iii)] an ``aggressive'' scenario, which would require more fundamental advances in techniques and tools.
\end{itemize}

In addition, there is the ``ideal'' scenario, in which theory uncertainties are subdominant for all observables and are thus negligible.

It should also be mentioned that the current framework of \ac{EWPO}s may not be sufficient for the experimental precision at future $e^+e^-$ colliders, and the discussion of a more extensive parameterisation~\cite{Jadach:2019bye, Blondel:2018mad} or new approaches inspired by lessons learned at the LHC~\cite{frixione-talk}, is starting within the community. For the purpose of the current report, a structure based on EWPOs is still adequate for assessing the impact of theoretical uncertainties on the physics reach of the colliders, even though a new theoretical framework is needed for the actual analysis of future experimental data.

The following notation is used in the next subsections to count complete and partial results at different orders of perturbation theory:
\begin{itemize}
    \item {\boldmath $\alpha$} denotes a complete electroweak order, including both QED and weak corrections. As a numerical factor, we use $\alpha = 1/137.037$;
    \item {\boldmath $\at$} is defined as $\at=y_{\rm t}^2/(4\pi)$, where $y_{\rm t}$ is the top Yukawa coupling. In practice, $\mathcal{O}(\at)$ corrections are obtained through a large-mass expansion;
    \item {\boldmath $\af$} refers to electroweak corrections with closed fermion loops. These are often simpler to compute than full $\mathcal{O}(\alpha)$ but are the numerically dominant piece of the latter;
    \item {\boldmath $\alpha_s$} denotes a complete QCD order, where numerically we use $\alpha_s \equiv \alpha_s(m_Z)$.
\end{itemize}

In what follows, we describe the estimates for the theory uncertainties that enter in the measurement and interpretation of the EWPOs and HPOs for the above-mentioned theory scenarios. We separate those directly affecting the experimental determination of the different (pseudo-)observables, from those associated to the \ac{SM} prediction for such quantities, and the determination of the SM input parameters.

\subsubsection{Extraction of pseudo-observables}
\label{sec:POextr}

The theory error from the subtraction of non-resonant “background” contributions to $Z$-pole pseudo-observables is treated as follows. In the current state-of-the-art scenario, \ac{NLO} corrections are included, while the future projection assumes that a complete \ac{NNLO} calculation will become available. This approach deliberately neglects partial higher-order effects that are already known at present, such as the resummed running of $\alpha_{\rm em}$, in order to maintain a consistent perturbative counting.
The associated theoretical uncertainties are estimated using a staged procedure. 
\begin{enumerate}
    \item[a)] First, predictions are evaluated with the GRIFFIN framework~\cite{Chen:2022dow} both including and excluding NLO corrections to contributions beyond the leading $Z$-pole term, in order to isolate the size of the non-resonant NLO corrections. To obtain the current uncertainty, this difference is multiplied by the factor $\frac{\alpha_W}{\pi} n_f$, where $\alpha_W = g^2/(4\pi)$ and $n_f = 12$ denotes the number of charged fermion species. 
    \item[b)] The future uncertainty, due to missing NNNLO electroweak effects, is estimated by applying the same factor once more, corresponding to an additional perturbative order.
\end{enumerate}
For mixed electroweak–QCD corrections, the treatment depends on the perturbative order considered. 
\begin{enumerate}
    \item[c)] At $\mathcal{O}(\alpha\alpha_s)$ (for the ``current'' scenario), the uncertainty is estimated using the known $\mathcal{O}(\af\alpha_s)$ corrections as a proxy. 
\item[d)] For the ``conservative'' future scenario, the leading missing contributions are $\mathcal{O}(\alpha^2\alpha_s)$, where the result obtained in (a) is rescaled by the relative factor $\alpha\alpha_s/\alpha_W$, and $\mathcal{O}(\alpha\alpha_s^2)$, which is estimated by multiplying the $\mathcal{O}(\alpha_f\alpha_s)$ contribution by the factor $\alpha_s/\pi \times 2C_A$.
\end{enumerate}
At present, no projections are available for the aggressive scenario. Consequently, this source of uncertainty is neglected in that case. The resulting uncertainty estimates are collected in Table~\ref{tab:therrZ1}.

\begin{table}[htbp]
	\centering
  \begin{tabular}{c c c c }
  \toprule
Observable & Current & Conservative & Aggressive \\
\midrule
\midrule
$\Gamma_Z$ (MeV) & $0.11$ & $0.016$ & --- \\[0.1cm]
\hline
&&&\\[-0.35cm]
$R_\ell$ ($10^{-3}$)& $2.1$ & $0.27$ & --- \\[0.1cm]
\hline
&&&\\[-0.35cm]
$R_b$ ($10^{-4}$) & $0.015$ & $0.0023$ & --- \\[0.1cm]
\hline
&&&\\[-0.35cm]
$R_c$ ($10^{-4}$) & $0.023$ & $0.0034$ & --- \\[0.1cm]
\hline
&&&\\[-0.35cm]
$\sigma_{\rm had}$ (pb) & $5$ & $0.8$ & --- \\[0.1cm]
\hline
&&&\\[-0.35cm]
$A_{\rm FB}^\ell$ ($10^{-4}$) & $3.4$ & $0.42$ & --- \\[0.1cm]
\hline
&&&\\[-0.35cm]
$A_{\rm FB}^b$ ($10^{-4}$) & $1.1$ & $0.22$ & --- \\[0.1cm]
\hline
&&&\\[-0.35cm]
$A_{\rm FB}^c$ ($10^{-4}$) & $0.8$ & $0.16$ & --- \\[0.1cm]
\hline
&&&\\[-0.35cm]
$A_{\rm LR}$ ($10^{-4}$) & $1.4$ & $0.29$ & --- \\[0.1cm]
\hline
&&&\\[-0.35cm]
$P_\tau$ ($10^{-4}$) & $3.0$ & $0.4$ & --- \\
\bottomrule
\end{tabular}
 	\caption{Current and projected theory uncertainties for the extraction of $Z$-pole POs at $e^+e^-$ colliders, due to background subtraction.}
    \label{tab:therrZ1}
\end{table}

The estimates of the uncertainties associated with QED \ac{MC} modeling, in particular those related to multi-photon radiation and fermion–pair creation, are based on the analysis presented in Ref.~\cite{Jadach:2019bye}, which provides a benchmark assessment of the size of these effects for $Z$-pole observables.
At present, no dedicated uncertainty estimates are available for the observables $R_{b,c}$, $A_{\rm FB}^{b,c}$, and $P_\tau$. For observables involving bottom and charm final states, the corresponding uncertainties are nevertheless expected to be smaller than those for the leptonic quantities $R_\ell$ and $A_{\rm FB}^\ell$. This expectation is motivated by the fact that the event selection for heavy-flavor final states is largely inclusive with respect to photon radiation in the final state. However, in the absence of a quantitative assessment, the degree of this reduction cannot currently be determined reliably.
As in the case of other theory-systematic effects, no projections are available for the aggressive scenario. This source of uncertainty is therefore neglected in that scenario. The resulting uncertainty estimates are summarized in Table~\ref{tab:therrZ2}.

\begin{table}[htbp]
	\centering
  \begin{tabular}{ c c c l }
  \toprule
Observable & Current & Conservative & Assumed improvements for ``Conservative'' \\
\midrule\midrule
$m_Z$ (MeV) & $0.3$ & $0.03$ & Fermion pairs \\
\hline
&&\\[-0.35cm]
$\Gamma_Z$ (MeV) & $0.2$ & $0.03$ & Fermion pairs \\
\hline
&&\\[-0.35cm]
$R_\ell$ ($10^{-3}$)& $12$ & $0.3$ & Fermion pairs; FSR+IFI beyond ${\cal O}(\alpha)$ \\
\hline
&&\\[-0.35cm]
$\sigma_{\rm had}$ (pb) & $25$ & $1.5$ & Fermion pairs; IFI beyond ${\cal O}(\alpha)$ \\
\hline
&&\\[-0.35cm]
$A_{\rm FB}^\ell$ ($10^{-4}$) & $5$ & $0.09$ & Full ${\cal O}(\alpha^2)$ QED plus log-enhanced ${\cal O}(\alpha^3)$ \\
\hline
&&&\\[-0.35cm]
$m_W$ (MeV) & $3$ & $0.3$ & \parbox{2.5in}{For $m_W$ from threshold scan:\newline
${\cal O}(\alpha^2)$ QED corrections for W production and decay; soft photon exponentiation for photons emitted from W bosons; improved ISR}\\
\bottomrule
\end{tabular}
 	\caption{Current and projected theory uncertainties for the extraction of $Z$-pole POs at $e^+e^-$ colliders, due QED modeling. Estimates for the aggressive scenario are unavailable at this time, and thus this source of uncertainty is ignored in this scenario.}
    \label{tab:therrZ2}
\end{table}

Uncertainties associated with QCD Monte Carlo (\ac{MC}) modeling, in particular those arising from hadronization and gluon splitting, are treated following the studies in Refs.~\cite{AlcarazMaestre:2020fmp,ALEPH:2005ab,Belloni:2022due}. For Z-boson branching-ratio observables, the current uncertainties are dominated by gluon-splitting contributions, which in principle can be reduced through improved perturbative calculations. In contrast, non-perturbative hadronization uncertainties originate from the limited precision of the data used for generator tuning and from the intrinsic modeling assumptions implemented in \ac{MC} generators. These non-perturbative effects can be probed, for example, by comparing predictions obtained with different hadronization models, such as string fragmentation and cluster fragmentation.
Based on these considerations, an overall improvement of the theory uncertainty on branching ratios by a factor of 10 in the conservative scenario and by a factor of 50 in the aggressive scenario, relative to the LEP estimates in Ref.~\cite{ALEPH:2005ab}, is assumed. 

For the forward–backward asymmetry $A_{\rm FB}$, it was pointed out in Ref.~\cite{AlcarazMaestre:2020fmp} that an acolinearity cut of 0.3 significantly reduces the sensitivity to theory modeling uncertainties, and this cut is assumed for all uncertainty estimates here. Compared to the analysis of Ref.~\cite{AlcarazMaestre:2020fmp}, for the conservative scenario it is assumed that non-perturbative hadronization uncertainties are reduced by a factor 5, and perturbative uncertainties reduced to $10^{-4}$ (as a relative factor) due to inclusion of NNLO QCD corrections (the techniques for calculating the latter are already available in~\cite{Bernreuther:2016ccf}). In the aggressive scenario, the perturbative treatment is extended to N$^3$LO QCD, and a factor of 50 improvement in non-perturbative effects (compared to today) is assumed.
The uncertainty associated with the N$^3$LO QCD corrections is estimated by rescaling the NNLO uncertainty by the factor $\alpha_s/\pi \times 2C_A$. Achieving a factor of 50 reduction in non-perturbative hadronization uncertainties would require not only extremely precise tuning data but also substantial conceptual and technical advances in hadronization modeling and simulation tools. Such a level of improvement is not expected to be achievable with the current string or cluster fragmentation models implemented in generators such as \textsc{Pythia} and \textsc{Herwig}~\cite{Andersson:1983ia,Bierlich:2022pfr,Webber:1983if,Bellm:2025pcw}.
The resulting uncertainty estimates are summarized in Table~\ref{tab:therrZ3}.

\begin{table}[htbp]
	\centering
  \begin{tabular}{c c c c }
  \toprule
Observable & Current & Conservative & Aggressive \\
\midrule\midrule
$R_b$ ($10^{-4}$) & $4.4$ & $0.44$ & $0.09$\\
\hline
&&&\\[-0.35cm]
$R_c$ ($10^{-4}$) & $17$ & $1.7$ & $0.34$ \\
\hline
&&&\\[-0.35cm]
$A_{\rm FB}^b$ ($10^{-4}$) & $1.0$ & $0.22$ & $0.028$ \\
\hline
&&&\\[-0.35cm]
$A_{\rm FB}^c$ ($10^{-4}$) & $0.75$ & $0.16$ & $0.021$ \\
\bottomrule
\end{tabular}
 	\caption{Current and projected theory uncertainties for the extraction of $Z$-pole POs at $e^+e^-$ colliders, due to QCD modeling.}
    \label{tab:therrZ3}
\end{table}

\subsubsection{Predictions for pseudo-observables}

Precise \ac{SM} predictions for electroweak and Higgs pseudo-observables (EWPOs/HPOs) are essential for comparison with the values extracted from data at future $e^+e^-$ colliders, and thereby for probing potential effects of physics beyond the Standard Model (BSM). In this context, the treatment of theoretical uncertainties for Higgs decay observables follows the methodology and assumptions outlined in Refs.~\cite{Lepage:2014fla,Freitas:2019bre}, with the decay $H\to Z\gamma$ treated according to Ref.~\cite{Chen:2024vyn}.
The size of missing higher-order corrections is estimated on the basis of the available partial perturbative results, combined with a geometric-series scaling to extrapolate the impact of yet-unknown contributions (see Eq.~\eqref{eq:geom}). 

\paragraph{Higgs boson decays:}
Table~\ref{tab:therrH} shows the estimated uncertainty of the currently available predictions for Higgs decays and future projections in two scenarios. Also shown are the perturbative orders that are currently known (see Refs.~\cite{Lepage:2014fla,Freitas:2019bre} and references therein) or assumed to be available in the future. Additional improvements for the aggressive scenario are only considered for decay channels where the conservative theory error estimate is not already sufficiently smaller than the projected experimental precision of future Higgs factories.
For the decays $H\to WW^*$ and $H\to ZZ^*$, reliable predictions require a calculation of the full process $H\to 4f$, including the decays of the intermediate gauge bosons, in order to correctly capture off-shell effects and interference contributions.
In the case of $H\to gg$, the dominant higher-order QCD corrections are computed in the infinite-$m_t$ limit within an effective field theory framework. Finite top-mass effects are known to be relatively modest, and corrections up to $\mathcal{O}(\alpha_s^2)$ are expected to be sufficient for the precision requirements considered here. For the loop-induced decays $H\to gg$, $H\to \gamma\gamma$, and $H\to Z\gamma$, electroweak corrections must be evaluated with the full top-mass dependence. In practice, this implies that the computation of the $n$th-order electroweak correction requires the evaluation of $(n+1)$-loop amplitudes.
For the decay $H\to Z\gamma$, the current SM predictions~\cite{Chen:2024vyn} are already adequate for the precision goals of future $e^+e^-$ colliders. For the FCC-hh, however, further theoretical improvements would be desirable, but a quantitative assessment of how much precision could be gained by extending the perturbative calculation to higher orders requires additional dedicated studies, and thus it is not considered in our scenarios.

\begin{table}[htbp]
	\centering
    {\small  % Use small to fit the table in the width of the document
  \begin{tabular}{ c l c l c l c l }
  \toprule
& & \multicolumn{2}{c}{\cellcolorCurr Current} & \multicolumn{2}{c}{\cellcolorC Conservative} & \multicolumn{2}{c}{Aggressive} \\
%\cdashline{2-7}
& &\cellcolorCurr Error &\cellcolorCurr &\cellcolorC Error &\cellcolorC Add'l. & Error & Add'l. \\
& Process &\cellcolorCurr est.~(\%) &\cellcolorCurr Available orders &\cellcolorC est.~(\%) &\cellcolorC orders & est.~(\%) & orders \\
\midrule\midrule
\!\multirow{6}{*}{\vtext{Decay}}&$H \to bb/cc$ &\cellcolorCurr $<0.4$ &\cellcolorCurr $\as^4+\al+\at^2+\at\as$ &\cellcolorC $\phantom{<~}0.2$ &\cellcolorC $\al^2+\al\as$ & $0.1$ & $\as^5$ \\
\cline{2-8}
&$H \to \tau\tau/\mu\mu$ &\cellcolorCurr $<0.3$ &\cellcolorCurr $\al+\at^2+\at\as$ &\cellcolorC $<0.1$ &\cellcolorC $\al^2+\al\as$ & & \\
\cline{2-8}
&$H \to WW^*\!/ZZ^*\!\!$ &\cellcolorCurr $\phantom{<~}0.5$ &\cellcolorCurr $\as+\al+\at^2+\at\as$ &\cellcolorC $\phantom{<~}0.3$ &\cellcolorC $\as^2$ & & \\
\cline{2-8}
&$H \to gg$ &\cellcolorCurr $\phantom{<~}3.2$ &\cellcolorCurr $\as^3+\al$ &\cellcolorC $\phantom{<~}1.0$ &\cellcolorC $\as^4$ & $0.5$ & $\as^5+\al\as$\! \\
\cline{2-8}
&$H \to \gamma\gamma$ &\cellcolorCurr $<1.0$ &\cellcolorCurr $\as^2+\al$ &\cellcolorC $<1.0$ &\cellcolorC & $0.4$ & $\al\as$\\
\cline{2-8}
&$H \to Z\gamma$ &\cellcolorCurr $\phantom{<~}1.5$ &\cellcolorCurr $\as+\al$ &\cellcolorC $\phantom{<~}1.5$ &\cellcolorC & & \\
\bottomrule
\toprule
\!\multirow{2}{*}{\vtext{ Prod.}}&$e^+e^- \to ZH$ &\cellcolorCurr$\phantom{<~}0.3$ &\cellcolorCurr $\al+\al\as+\al\af$ &\cellcolorC $<0.1$ &\cellcolorC $\al^2$ & & \\
\cline{2-8}
&$e^+e^- \to \nu\bar\nu H$ &\cellcolorCurr $\sim 1$ &\cellcolorCurr $\al$ &\cellcolorC $\sim 0.1$ &\cellcolorC $\al\as+\al^2$ & & \\
\bottomrule
\end{tabular}
 	\caption{Theory uncertainties (in percent) for the SM predictions of partial Higgs decay widths and production cross-sections (for $\sqrt{s} \sim 230$\! --\! $550$~GeV).}
    \label{tab:therrH}
    }
\end{table}

\paragraph{Higgs boson production:}
The theoretical uncertainty estimates for Higgs production processes are defined as follows. For associated $HZ$ production, the current error estimate is taken from Ref.~\cite{Freitas:2023iyx}, which is mainly driven by the missing NNLO electroweak corrections without closed fermion loops. For Higgs production via $WW$ fusion at $e^+e^-$ colliders, only NLO corrections are currently known~\cite{Denner:2003yg}.

Future uncertainty estimates are obtained using a prefactor-based scaling approach. Missing higher-order electroweak corrections of $\mathcal{O}(\alpha^3)$ are estimated by rescaling the $\mathcal{O}(\alpha^2)$ uncertainty according to
\begin{align}
\delta(\alpha^3) &= \delta(\alpha^2) \times \frac{\alpha_W}{\pi} n_f,
\qquad \alpha_W = \frac{g^2}{4\pi}, \; n_f = 12 \;\text{(number of charged fermion species)}. \notag
\end{align}
Mixed electroweak–QCD corrections are treated analogously. For $\mathcal{O}(\alpha^2\alpha_s)$ contributions, the uncertainty is estimated as
\begin{align}
\delta(\alpha^2\alpha_s) &= \delta(\alpha^2) \times \frac{\alpha_s}{\pi} 2C_F
\qquad \parbox[t]{4in}{(with a heuristic factor of 2 to account for diagram combinatorics),} \notag
\end{align}
while higher-order QCD effects at $\mathcal{O}(\alpha\alpha_s^2)$ are obtained by rescaling the $\mathcal{O}(\alpha\alpha_s)$ uncertainty as
\begin{align}
\delta(\alpha\alpha_s^2) &= \delta(\alpha\alpha_s) \times \frac{\alpha_s}{\pi} 2C_A,
\qquad \parbox[t]{4in}{(with a heuristic factor of 2 to account for diagram combinatorics).} \notag
\end{align}
For the $WW$ fusion process, we simply adopt the same relative uncertainty estimates as for the $HZ$ process.
The resulting theoretical uncertainty estimates for Higgs production observables are summarized in Table~\ref{tab:therrH}.

The numbers shown in this table are assumed to be valid for center-of-mass energies up to 500~GeV. However, the vector-boson fusion process, $\ell^+\ell^- \to \nu_\ell\bar\nu_\ell H$, can be quite important at higher energies, such as (multi-)TeV linear $e^+e^-$ colliders and/or a muon collider. At these energies, fixed-order theory calculations will be dominated by electroweak Sudakov logarithms. Based on this observation, we estimate that the theory uncertainty will increase by a factor $\frac{\ln^2(\sqrt{s}/m_W)}{\ln^2((500~\!{\rm GeV})/m_W)}$ relative to the estimate for $\sqrt{s}=500$~GeV, \emph{i.e.} in the conservative scenario it is about 0.2\% at $\sqrt{s} = 1$~TeV, 0.4\% at 3~TeV, and 0.7\% at 10~TeV, respectively.

%\paragraph{$Z$ pole (incl.\ $W$-mass prediction):}
\paragraph{$Z$-pole observables and $W$-mass:}
The current theoretical uncertainty estimate is taken from Ref.~\cite{Awramik:2003rn,Dubovyk:2019szj}. The results in these papers include calculations at ${\cal O}(\alpha^2)$ and ${\cal O}(\alpha\alpha_s)$, as well as higher-order top-Yukawa–enhanced terms of ${\cal O}(\alpha_t\alpha_s^2)$, ${\cal O}(\alpha_t^2\alpha_s)$, ${\cal O}(\alpha_t^3)$, and ${\cal O}(\alpha_t\alpha_s^3)$. In addition, final-state QCD corrections are included up to ${\cal O}(\alpha_s^4)$. (See Ref.~\cite{Awramik:2003rn,Dubovyk:2019szj} for the complete list of references.) Not included are the three-loop corrections with a maximal number of fermion loops~\cite{Chen:2020xzx,Chen:2020xot}, but those are believed to have only a minor impact on the error estimate.

In the conservative future scenario, it is assumed that a number of further perturbative contributions become available beyond the currently known leading-$\alpha_t$ approximation. These include mixed electroweak–QCD terms at ${\cal O}(\alpha\alpha_s^2)$, ${\cal O}(\alpha^2\alpha_s)$, and ${\cal O}(\alpha_f^2\alpha)$, as well as higher-order top-Yukawa–enhanced corrections at ${\cal O}(\alpha_t^2\alpha_s^2)$, ${\cal O}(\alpha_t^3\alpha_s)$, and ${\cal O}(\alpha_t^4)$. Furthermore, final-state QCD corrections are assumed to be extended to ${\cal O}(\alpha_s^5)$.
For the aggressive scenario, additional progress beyond this level is envisaged. Full corrections beyond the leading-$\alpha_t$ approximation are assumed to be available at ${\cal O}(\alpha^3)$, ${\cal O}(\alpha^2\alpha_s^2)$, ${\cal O}(\alpha_f^2\alpha\alpha_s)$, and ${\cal O}(\alpha_f^2\alpha^2)$. Moreover, higher-order top-Yukawa–enhanced terms at ${\cal O}(\alpha_t\alpha_s^4)$ and ${\cal O}(\alpha_t^2\alpha_s^3)$ are included, and final-state QCD corrections are assumed to be known up to ${\cal O}(\alpha_s^6)$. Note that in both scenarios, at the highest loop orders we only consider corrections with some number of closed fermion loops, since these are expected to be numerically dominant and technically more manageable.
The projected uncertainties for both future scenarios are estimated using the extrapolation-of-perturbative-series method. The resulting uncertainty estimates are summarized in Table~\ref{tab:therrZ}.

\begin{table}[htbp]
	\centering
  \begin{tabular}{c c c c }
  \toprule
Observable & Current & Conservative & Aggressive \\
\midrule\midrule
$\Gamma_Z$ (MeV) & 0.4 & 0.08 & 0.016 \\
\hline
&&&\\[-0.35cm]
$R_\ell$ ($10^{-3}$) & 6 & 1.2 & 0.2 \\
\hline
&&&\\[-0.35cm]
$R_b$ ($10^{-4}$) & 1 & 0.2 & 0.035 \\
\hline
&&&\\[-0.35cm]
$R_c$ ($10^{-4}$) & 0.5 & 0.1 & 0.02 \\
\hline
&&&\\[-0.35cm]
$\sigma_{\rm had}$ (pb) & 6 & 1.6 & 0.3 \\
\hline
&&&\\[-0.35cm]
$\sin^2{\theta}_{\rm eff}$ ($10^{-5}$) & $4.5$ & $0.7$ & $0.06$ \\
\hline
&&&\\[-0.35cm]
$m_W$ (MeV) & $4.0$ & $1.0$ & $0.1$\\
\bottomrule
\end{tabular}
 	\caption{Current and projected theory uncertainties for the SM prediction of $Z$-pole POs and the $W$ mass.}
    \label{tab:therrZ}
\end{table}

\paragraph{W-boson decays and $WW$ production:}
The treatment of theoretical uncertainties for W-boson observables follows the same general strategy adopted for Higgs observables, combining the best available fixed-order calculations (see Ref.~\cite{dEnterria:2020cpv} and references therein) with proxy-based estimates for missing higher-order contributions.
For the leptonic partial decay width $W\to \ell\nu$, next-to-leading-order (NLO) corrections are currently known. The ${\cal O}(\alpha_f^2)$ corrections are comparatively straightforward to compute and have been explicitly evaluated for the present study. The latter result is used as a proxy for the size of the missing complete NNLO electroweak corrections.
For hadronic decays, $W\to qq'$, perturbative corrections are available at ${\cal O}(\alpha)$, ${\cal O}(\alpha\alpha_s)$, and ${\cal O}(\alpha_s^4)$. The uncertainty associated with missing higher-order QCD contributions is estimated following the methodology of Ref.~\cite{dEnterria:2020cpv}. However, in this channel the dominant theoretical uncertainty also arises from missing electroweak contributions beyond NLO. In the conservative future scenario, it is assumed that the full NNLO corrections become available, together with mixed electroweak–QCD contributions at ${\cal O}(\alpha\alpha_s^2)$ for $qq'$ final states.
Future uncertainties are estimated using the same prefactor-based scaling procedure employed elsewhere in this study. Missing electroweak orders are accounted for by multiplying the current uncertainty by the factor $\frac{\alpha_W}{\pi} n_f$, while missing QCD orders are estimated by applying the factor $\frac{\alpha_s}{\pi} C_A$.

For the ratio of hadronic to leptonic branching fractions, the leading radiative corrections largely cancel. As a consequence, the current theoretical uncertainty is dominated by missing mixed electroweak–QCD corrections at N$^3$LO, as discussed in Ref.~\cite{dEnterria:2020cpv}. Both the current and projected uncertainties for this observable are evaluated using the same scaling assumptions as for the individual partial widths.
For $e^+ e^- \to W^+W^-$ production, the current state of the art consists of NLO predictions including off-shell effects, with an associated uncertainty estimate provided in Ref.~\cite{Denner:2005es}. The conservative future scenario assumes the availability of full NNLO corrections for on-shell (double-pole) $W^+W^-$ production. The projected uncertainty is obtained by rescaling the current estimate using the same electroweak and QCD prefactors defined above.

The resulting projected uncertainties in the conservative scenario are sufficiently small that no separate aggressive scenario is introduced for W-boson observables. All numerical results are summarized in Table~\ref{tab:therrW}.

\begin{table}[htbp]
	\centering
  \begin{tabular}{l c c c }
  \toprule
Quantity & Current & Conservative & Aggressive\\
\midrule\midrule
$\Gamma_W^{\rm lep}$ (\%) & $0.1$ & $0.013$ & --- \\
\hline
&&&\\[-0.35cm]
$\Gamma_W^{\rm had}$ (\%) & $0.1$ & $0.015$ & --- \\
\hline
&&&\\[-0.35cm]
$\Gamma_W^{\rm had}/\Gamma_W^{\rm lep}$ (\%) & $0.015$ & $<0.01\phantom{00}$ & --- \\
\hline
&&&\\[-0.35cm]
$\sigma[e^+e^- \to W^+W^-]$ (\%) & $0.4$ & $0.07$ & --- \\
\bottomrule
\end{tabular}
 	\caption{Theory uncertainties (in percent) for the SM predictions of partial $W$-boson decay widths, ratios, and production cross-sections (for $\sqrt{s} \sim 230$\! --\! $550$~GeV).}
    \label{tab:therrW}
\end{table}

\paragraph{Fermion pair production:}
The evaluation of theoretical uncertainties for fermion-pair production follows the same strategy adopted for the subtraction of non-resonant backgrounds discussed in Section~\ref{sec:POextr}. Thus the current scenario includes NLO electroweak corrections together with fourth-order QCD corrections for light-quark $q\bar{q}$ final states ($q\neq t$). The (conservative) future projection assumes that the full NNLO corrections and third-order mixed electroweak corrections at $\mathcal{O}(\alpha^2\as)$ and $\mathcal{O}(\alpha\as^2)$ will become available.
The uncertainties are estimated by applying appropriate pre-factors as described in Section~\ref{sec:POextr}.
For light-quark final states, the uncertainty associated with missing fifth-order QCD corrections is negligible compared to the electroweak uncertainties and is therefore omitted.

In the case of $t\bar{t}$ production, next-to-next-to-next-to-leading-order (N$^3$LO) QCD corrections are available, and the corresponding uncertainty estimate is taken from the analysis presented in Ref.~\cite{Chen:2022vzo}. For the future conservative scenario, it is assumed that N$^4$LO QCD corrections become available. The associated uncertainty is then assumed to be further suppressed by a factor $\alpha_s/v_t$, where $v_t$ denotes the top-quark velocity and parametrizes the enhancement of radiative corrections in the vicinity of the production threshold.

For the aggressive scenario, no uncertainty projections are available, and thus the theory error for fermion pair processes is neglected in this scenario. The resulting uncertainty estimates are summarized in Table~\ref{tab:therrF1}.

\begin{table}[htbp]
	\centering
  \begin{tabular}{c c c c c }
  \toprule
& \multicolumn{2}{c}{$\sqrt{s} = 250$~GeV} & \multicolumn{2}{c}{$\sqrt{s} = 365$~GeV} \\
%\cline{2-5}
Quantity & Current & Conservative & Current & Conservative \\
\midrule\midrule
$\sigma[e^+e^- \to \ell^+\ell^-~\!]$ (\%) &
1.2 & 0.15 & 1.2 & 0.15 \\
\hline
&&&&\\[-0.35cm]
$\sigma[e^+e^- \to c\bar{c}~\!]$ (\%) & 0.8 & 0.1 & 0.7 & 0.09 \\
\hline
&&&&\\[-0.35cm]
$\sigma[e^+e^- \to b\bar{b}~\!]$ (\%) & 0.43 & 0.06 & 0.45 & 0.06 \\
\hline
&&&&\\[-0.35cm]
$\sigma[e^+e^- \to \mathrm{had.}~\!]$ (\%) & 0.6 & 0.08 & 0.6 & 0.08 \\
\hline
&&&&\\[-0.35cm]
$\sigma[e^+e^- \to t\bar{t}~\!]$ (\%) & --- & --- & 1.2 & 0.4 \\
&&&&\\[-0.4cm]
  \toprule
  \bottomrule
&&&&\\[-0.35cm]
$A_{\rm FB}^\ell$ (\%) & 0.4 & 0.05 & 0.35 & 0.045 \\
\hline
&&&&\\[-0.35cm]
$A_{\rm FB}^c$ (\%) & 0.47 & 0.06 & 0.64 & 0.08 \\
\hline
&&&&\\[-0.35cm]
$A_{\rm FB}^b$ (\%) & 0.25 & 0.03 & 0.3 & 0.04 \\
\bottomrule
\end{tabular}
 	\caption{Current and projected theory uncertainties for the extraction fermion pair production
    above the $Z$-pole at $e^+e^-$ colliders. ``had'' stands for all hadronic final states except top quarks.}
    \label{tab:therrF1}
\end{table}

\subsubsection{\ac{SM} parameters}

The precision physics program of future $e^+e^-$ colliders requires a much improved determination of other \ac{SM} parameters which are used as inputs for theory predictions. The extraction of these parameters from data  is also subject to theory uncertainties. The determinations of the masses of the top quark and the $W$ boson from pair production near threshold require predictions of the cross-section including off-shell and Coulomb scattering effects. They can be obtained in an integrated effective field theory framework. The estimates in Table~\ref{tab:therrSMpar} are based on Ref.~\cite{Freitas:2019bre,Defranchis:2025auz}. Estimates for the aggressive future theory scenario are currently not available, but further reduction of theoretical uncertainties will require fundamental advances in computational techniques. 

The most important source of uncertainty of the strong and electromagnetic couplings at the scale $m_Z$ stems from non-perturbative QCD, which could be reduced with advances in lattice calculations. The strong coupling could also be determined from the $Z$-pole ratio $R_\ell=\Gamma_{Z\to{\rm had}}/\Gamma_{Z\to \ell^+ \ell^-}$~\cite{Proceedings:2015eho}, but this requires the assumption that there are no contributions of BSM physics in this quantity,
so we are not considering this option as input for the global fits.

The electromagnetic coupling $\alpha(m_Z)$ could also be directly determined from measurements near the $Z$-pole. To achieve a competitive precision, the high luminosity of a circular $e^+e^-$ collider would be needed for that purpose. Ref.~\cite{Janot:2015gjr} uses measurements of $A_{\rm FB}^\mu$ at two center-of-mass energies $\sqrt{s_{1,2}} \sim m_Z \pm$ few GeV, which help to reduce the impact of systematic uncertainties. At these energies, the matrix elements for $e^+e^- \to \mu^+\mu^-$ receive significant interference between photon and $Z$ s-channel amplitudes, where the former depends on $\alpha(s)$, while the latter is assumed to be well constrained from measurements at $\sqrt{s} = m_Z$, thus allowing for the extraction of $\alpha(s_{1,2})$. However, there are radiative corrections to $e^+e^- \to \mu^+\mu^-$ that are not captured by $Z$-boson effective couplings and/or $\alpha(s)$, in particular box contributions, that need to be supplied from theory and thus have an intrinsic theory uncertainty~\cite{Freitas:2019bre}.

Ref.~\cite{Riembau:2025ppc} proposes to use measurements of the differential cross-section of $e^+ e^- \to e^+ e^-$ at $\sqrt{s}=m_Z$, which are dominated by s-channel $Z$-boson exchange at large scattering angles, and t-channel photon exchange at small angles. The latter allows for the extraction of $\alpha(t)$ for space-like $t<0$. Theory uncertainties enter through the need to translate $\alpha(t)$ to the time-like $\alpha(m_Z)$, and through radiative corrections to the process $e^+ e^- \to e^+ e^-$, in particular vertex and box contributions, whose impact on this analysis has not been evaluated so far.

\begin{table}[htbp]
	\centering
  \begin{tabular}{l c c c }
  \toprule
Quantity & Current & Conservative & Aggressive\\
\midrule\midrule
$m_t$ (MeV) & $35$ & $25$ & --- \\
\hline
$m_W$ (MeV) & $3$ & $0.6$ & --- \\
\hline
$\as(m_Z)$ \, ($10^{-3}$) & 1 & $0.3$ & $0.1$ \\
\hline
$\alpha(m_Z)/\alpha(0)$ \, ($10^{-4}$) & $<1\phantom{.~\!0}$ & $<0.5\phantom{.~\!0}$ & $0.1$\! --\! $0.3$ \\
\bottomrule
\end{tabular}
 	\caption{Impact of theory errors on the determination of various SM parameters.}
    \label{tab:therrSMpar}
\end{table}

\begin{figure}[ht!]
\begin{center}
\includegraphics[width=0.78\textwidth]{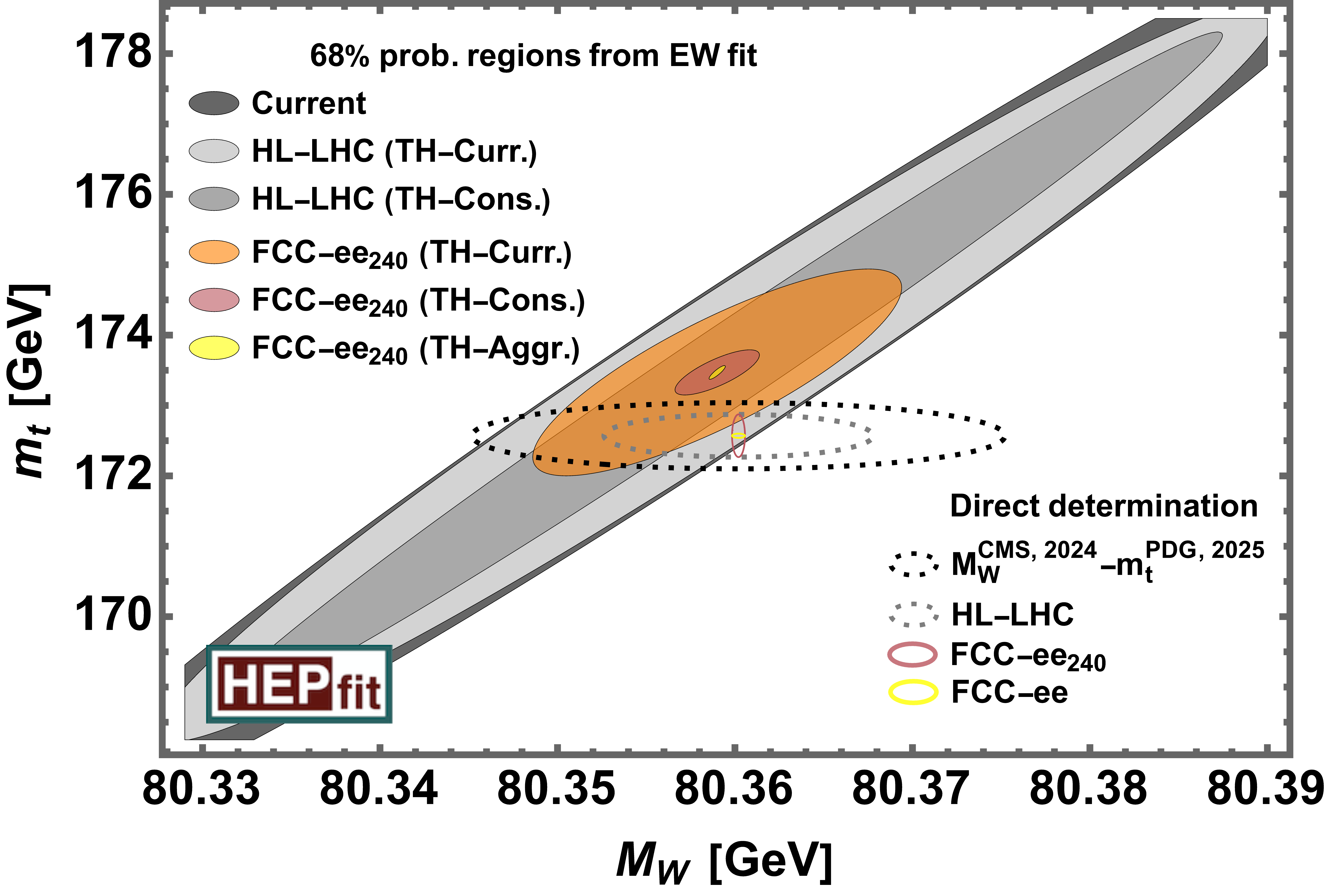}\\
\vspace{0.75cm}
\hspace{-1.1cm}\includegraphics[width=0.8\textwidth]{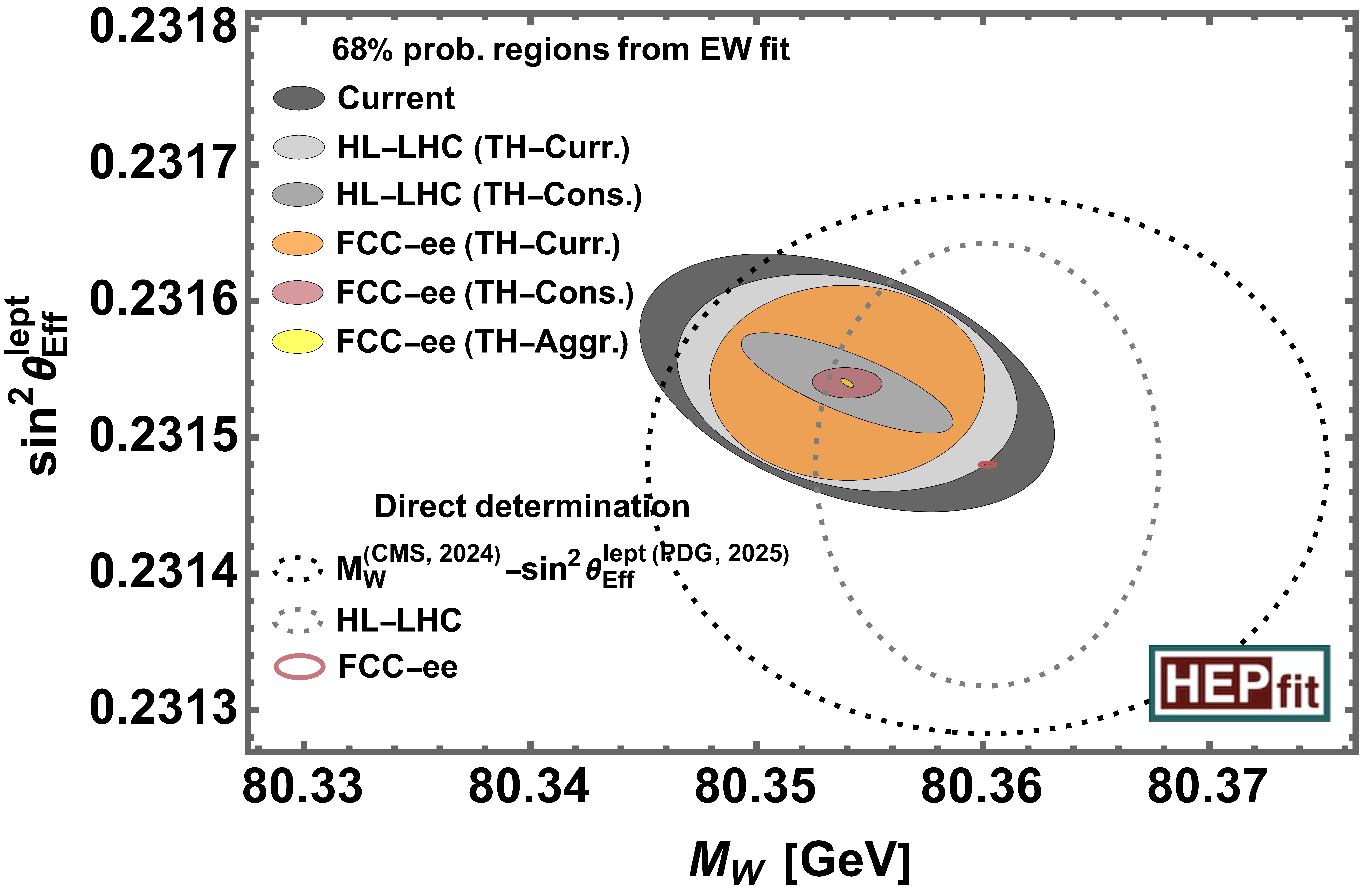}
\end{center}
\caption{
Top: Comparison of the indirect and direct mass determination of the $W$ and top quark for current data, the HL-LHC and the FCC--ee, and different assumptions for the size of theory uncertainties. 
Bottom: The same for the $W$ mass and the leptonic weak mixing angle.}
\label{fig:ew:SM}
\end{figure}

To conclude this section, we show the impact of the different assumptions for theory uncertainties in EWPOs in Figure~\ref{fig:ew:SM} for the FCC--ee, compared to the HL-LHC and current EW precision data. 
(This is similar to the top panel in Figure 3.3 in~\cite{deBlas:2025PhysicsBriefingBook}.)
Aside from the results for the aggressive and conservative scenarios, we also display the impact of not improving the theory calculations with respect to the current ones.~\footnote{In this regard, the uncertainty in the SM prediction of the $W$ mass is taken to be 4 MeV, as in Table~\ref{tab:therrZ}. However, this estimate ought to be revisited in view of the recent calculation of the third order mixed EW-QCD corrections to the SM calculation of $m_W$ from the muon decay in Ref.~\cite{Dubovyk:2026nhx}.}

% -------------------------------------

\section{Additional information on PPG EW WG project comparisons}

In the next subsections we describe some details of the PPG EW WG comparisons. These were obtained from a series of fits to the projected EW, Higgs and Top-quark measurements for different theory frameworks: (a) the formalism of electroweak oblique corrections~\cite{Peskin:1990zt,Peskin:1991sw}, for the assessment of precision EW measurements; (b) the kappa framework~\cite{LHCHiggsCrossSectionWorkingGroup:2012nn,LHCHiggsCrossSectionWorkingGroup:2013rie}, for Higgs precision; and (c) the \ac{SMEFT} framework, see e.g.~\cite{Brivio:2017vri,Isidori:2023pyp,Aebischer:2025qhh}, for a global interpretation of the EW, Higgs and Top observables. In the following, we provide a short summary of the frameworks, the numerical values of the results presented in Ref~\cite{deBlas:2025PhysicsBriefingBook}, and some additional results which were not included in the Physics Briefing Book. We also provide additional information on the inputs entering in these studies, and some technical details and references of the tools used for the fits.

All fits are performed assuming the expected HL-LHC dataset and precision as the baseline scenario. This HL-LHC scenario, also described below, provides the reference against which the sensitivity of each future collider is assessed.

\subsection{Fit and statistical framework}

All the results of the EW WG comparisons were obtained using the {\tt HEPfit} code~\cite{DeBlas:2019ehy,Aebischer:2023nnv}, where the last version can be obtained in the {\tt GitHub} repository~\cite{hepfitsite}. The studies were performed using a Bayesian statistical analysis, performing a Markov Chain Monte Carlo (MCMC) sampling the posterior density function (pdf) of the different fitting parameters (e.g.~the Higgs coupling modifier, $\kappa_i$, in the kappa framework, or the Wilson coefficients $C_i$ in the SMEFT comparison). From the MCMC samples we obtain statistical information on the parameters, namely 68\% or 95\% Highest Posterior Density intervals, correlations, etc. In particular, for most of the results we quote the {\em 68$\%$ probability sensitivity} for a given parameter, defined as half of the size of the corresponding 68$\%$ probability interval for that quantity. This corresponds to the 1$\sigma$ uncertainty in the gaussian case.

Inside {\tt HEPfit} the following model classes are used for the different comparisons, all within the {\tt NewPhysics} project:~\footnote{Some of the results presented in~\cite{deBlas:2025PhysicsBriefingBook}, namely those in the top panel in Figure 3.3, were obtained under the SM hypothesis and can be reproduced using the {\tt StandardModel} class of the project with the same name.} 
(1) the class {\tt NPSTU} is used for the study in terms of oblique parameters in section~\ref{sec:STpars}; 
(2) the class {\tt HiggsKigen} for the comparisons of Higgs precision in the kappa framework, see section~\ref{sec:Kappa}; 
and (3) the model classes {\tt NPSMEFTd6General} and {\tt NPd6SILH} for the comparisons in the dimension-six SMEFT framework in section~\ref{sec:EFT}.~\footnote{The class {\tt NPd6SILH} is only used for the fit results presented in Figure 3.9 in Ref.~\cite{deBlas:2025PhysicsBriefingBook}. All the other EFT results in the EW chapter of~\cite{deBlas:2025PhysicsBriefingBook} were obtained using the general SMEFT class {\tt NPSMEFTd6General}.}
The latter also performs the renormalization group evolution of the SMEFT Wilson coefficients, using the {\tt RGEsolver} library~\cite{DiNoi:2022ejg}. 

In the calculation of the likelihood we  assume the observables are measured with the uncertainties corresponding to the projections from each collider, whereas for the central values we always assume the SM prediction.

\subsection{Inputs used in the fits}

The baseline projections used in the fits are based on the original submissions from the projects to the ESPP. In several cases, however, those inputs are obtained from studies performed under different assumptions that are not intrinsic to the physics potential of the colliders, such as the use of fast versus full detector simulation, the treatment of theoretical systematic uncertainties, and the level of analysis optimization. To improve the consistency of the comparison, we have standardized these assumptions wherever possible. In the following, we summarize the Higgs, electroweak, and top-quark inputs used in the fits, with emphasis on the modifications made to the $e^+e^-$ collider inputs to harmonize these assumptions across studies. 
\paragraph{Higgs-boson observables:}
For the observables common to all future $e^+ e^-$ Higgs factories, namely those that could be measured at energies up to $\sqrt{s}\sim 350$--380 GeV, the projections for all colliders were obtained from a common set of inputs where, for each channel, the best-performing result among the different colliders was taken as the baseline.\footnote{The exception is the $H\to {\rm invisible}$ channel, where we take an average of the FCC--ee and LCF results. } This was then extrapolated to the different projects, taking into account the differences in beam polarization, running energy and integrated luminosity. The inputs for energies above $\sqrt{s}=500$ GeV were used as provided in the original submission documents and the references therein. For di-Higgs production at the LCF at energies above 500 GeV, we use the projections from~\cite{Berggren:2025fpw}. For the case of CLIC, the di-Higgs projections have not been revisited since the work in Ref.~\cite{Roloff:2019crr} and we still use those in this study.

The inputs for other collider types ($pp$, $ep$, or $\mu^+\mu^-$) were also taken directly from the original submission documents and the references therein. For LHeC, we use all the available information in Ref.~\cite{LHeC:2020van}, including both charged- and neutral-current Higgs boson production in deep-inelastic scattering. For the 10-TeV muon collider, we use the projections for single-Higgs measurements derived assuming forward muon tagging, as documented in Ref.~\cite{Forslund:2022xjq,Ruhdorfer:2024dgz}. The muon collider sensitivity to the Higgs self-coupling is derived from an updated version of the study in Ref.~\cite{Buttazzo:2020uzc}. Finally, for the FCC--hh, aside from the information in the ESPP input Ref.~\cite{Selvaggi:2025kmd}, which contains the projections for single- and double-Higgs production, we also use the projections for differential observables at high energies for $WH$ and $ttH$ from Ref.~\cite{mazzeo_2026_3p8sa-33196}.

\paragraph{Electroweak precision observables:}

For the EWPOs, the main difference between the submitted inputs lies in the treatment of theoretical systematic uncertainties. The FCC--ee projections factor out these uncertainties, together with possible detector systematics, while the LC projections include them explicitly. The final set of inputs was obtained in consultation with both projects by defining a common set of projections in which these effects were removed. We also revised the estimates of the experimental systematic uncertainties where they are expected to scale with the statistical precision.
The projections for the LEP3 EWPOs were obtained from the FCC--ee ones, adjusting for the different luminosity and beam polarization.\footnote{We note that, for some observables like $m_W$, the systematics from beam-energy calibration at LEP3 cannot be directly extrapolated from the FCC--ee ones, due to the lack of resonant depolarization at the former at the $WW$ threshold energies. A separate estimate based on the use of $Z\gamma$ for the calibration was used in that case.} These results are summarized in Ref.~\cite{de_blas_2026_wbvaf-gsk41} and the LC submission~\cite{sub40} has been updated to reflect these changes.\footnote{Due to a typo in the preliminary version of Ref.~\cite{sub40} used to prepare the comparisons, the LCF precision on the $W$-boson leptonic branching fractions was set to the same value as the FCC--ee:~$(\pm 0.13_{\rm Stat} \pm 0.1_{\rm Sys})\cdot 10^{-4} $. 
} 
The theoretical systematics were then added in quadrature to the resulting experimental uncertainties, under the {\it aggressive} and {\it conservative} assumptions detailed in the previous section. 

Running at the $Z$ pole or $WW$ threshold is not part of the physics programme of muon colliders. For the 10 TeV muon collider, information about the fermionic EW interactions can be obtained by studying, e.g., VBF production of fermion pairs, and this is included in our fits~\cite{InternationalMuonCollider:2025sys}. For LHeC, information about the EW interactions can be derived from studies of deep-inelastic scattering. We include the SMEFT results of Ref.~\cite{Bissolotti:2023vdw}. 

\paragraph{Other electroweak observables:} In addition to the $Z$-pole and $W$-boson measurements entering the EWPO, we also include projections for diboson production and two-to-two fermion processes. In the case of $e^+ e^- \to W^+ W^-$, the projections for all $e^+e^-$ colliders were obtained using the optimal observable method as described in Ref.~\cite{DeBlas:2019qco}. We use the same final states as in the Snowmass study in Ref.~\cite{deBlas:2022ofj}, rescaling the statistical uncertainty to the projected luminosity submitted by the projects. Similarly, we use the $e^+e^- \to f\bar{f}$ projections for the determination of the total cross section and forward-backward asymmetries in Ref.~\cite{deBlas:2022ofj}, also appropriately rescaled to account for the different luminosity. For several of the hadronic final states, more detailed projections for the LCF project are discussed in Ref.~\cite{LinearColliderVision:2025hlt}, and are used here. The same applies to the inputs for the FCC--ee, where we use the projections for measurements above the $Z$ resonance in Ref.~\cite{blondel_2025_3rj9z-94994}, complemented with the information from Ref.~\cite{deBlas:2022ofj} for the $e^+e^-$ final state. Finally, for the case of LEP3, the projections are again obtained from the FCC--ee ones. In this case, we assume the systematic uncertainties are the same as those from FCC--ee and only account the different luminosity of the runs at 161 GeV and 230 GeV. 

For the 10 TeV muon collider, projections for several EW di-fermion and di-boson production channels are included in the fits~\cite{Chen:2022msz}. For FCC-hh, differential information on $W^+W^-$ and di-fermion production is available~\cite{selvaggi_2025_84fpn-2qe63,FCChhDifferential}. Finally, for the LHeC, we use the projected constraints on the anomalous triple-gauge couplings directly in the fit~\cite{LHeC:2020van}.

\paragraph{Top observables:}
The projections for the EFT constraints from top pair production at the different $e^+ e^-$ colliders have been obtained via the statistical optimal observable method described in  Ref.~\cite{Durieux:2018tev}, adapted to the energies and luminosities of each project in the comparison. At the 10 TeV muon collider, the sensitivity to top interactions in our fit comes from the study of $\mu^+\mu^-\to t\bar{t}$ production at high-energies, as well as VBF production of top pairs~\cite{Chen:2022msz,InternationalMuonCollider:2025sys}. At the FCC--hh differential information for the following Top-processes is included: $tt$, $tb$, $tttt$ and $ttZ$~\cite{selvaggi_2025_84fpn-2qe63,Beriet:2026dyk}.   \\

\paragraph{HL-LHC baseline:}
For the EWPO, the baseline fit includes the LEP/SLC measurements, together with the projected HL-LHC uncertainties on the $W$ boson and top quark masses, which we take to be 5 MeV and 200 MeV, respectively.~\footnote{For the SM and oblique parameter interpretations we also assume the HL-LHC determination of the effective weak mixing angle will reach a precision of $15\times 10^{-5}$~\cite{Azzi:2019yne}.} For the Higgs sector, we include the projections for single-Higgs measurements, updating the information from Ref.~\cite{deBlas:2019rxi} using the results in Ref.~\cite{ATLAS:2025eii}. We validated this updated input by performing the same $\kappa$-fit reported in Figure 1 in Ref.~\cite{ATLAS:2025eii} and comparing our results with the ones in that reference. 
We also include the updated precision for the Higgs self-coupling from projected Higgs pair production measurements.
Ref.~\cite{ATLAS:2025eii} does not include projections for diboson measurements, which are particularly relevant for the SMEFT interpretation of Higgs observables. We therefore include the projections from the theoretical study in Ref.~\cite{Grojean:2018dqj} (also used in Ref.~\cite{deBlas:2022ofj}). For the top-quark sector, we use the extrapolation of the effective field theory analysis of Run 2 data presented in Ref.~\cite{CMS:2023xyc} to the HL-LHC, assuming a total integrated luminosity of 3 ab$^{-1}$ for both ATLAS and CMS, 
as presented in Ref.~\cite{Collaboration:2938605}.~\footnote{While there are currently no official projections for the precision of measurements of the Drell-Yan process at HL-LHC, for the results in the section {\it Precision from Energy} in Chapter 3 in \cite{deBlas:2025PhysicsBriefingBook} we use the results from the theory study in~\cite{Farina:2016rws}.} 
The EFT results from that study are then combined with the remaining inputs in our fit.

\subsection{Electroweak constraints on Oblique parameters\label{sec:STpars}}

\begin{figure}[ht!]
\begin{center}
\includegraphics[width=0.7\textwidth]{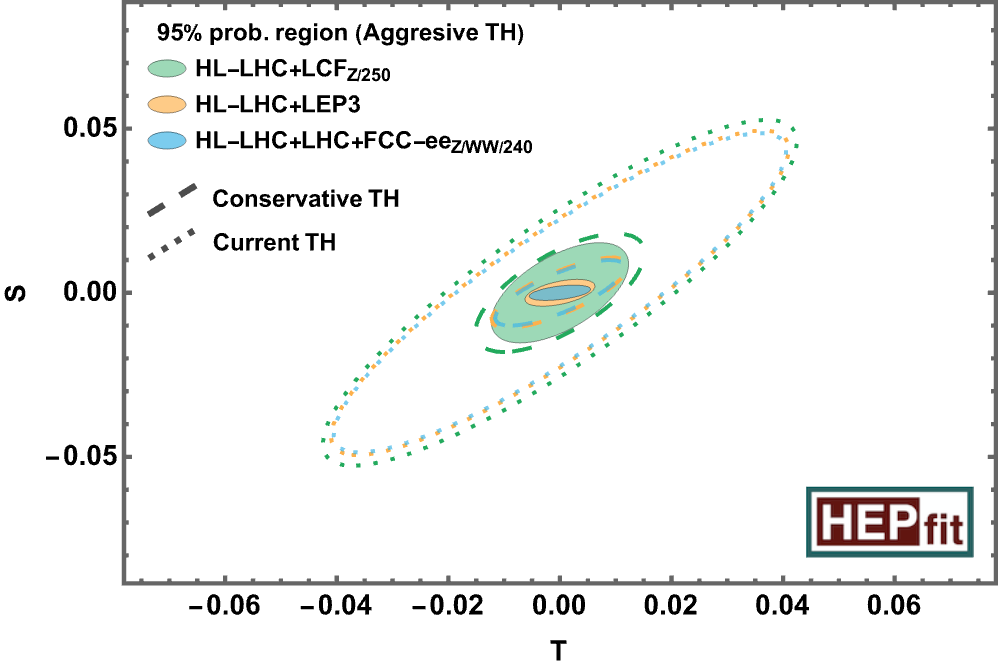}\\
\vspace{0.75cm}
\includegraphics[width=0.7\textwidth]{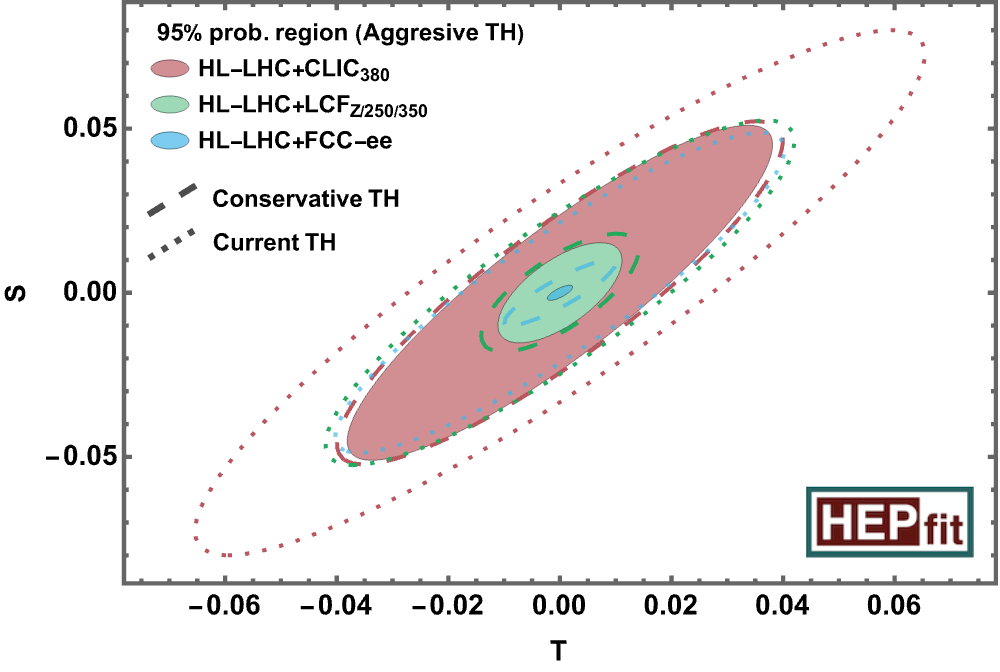}
\end{center}
\caption{
Top: Comparison of the constraints on the $S$ and $T$ parameters for $e^+ e^-$ colliders operating below the $t\bar{t}$ threshold. We compare the results assuming the aggressive (solid regions) and conservative (dashed lines) scenarios for theory uncertainties, and also the impact of not improving the theory calculations with respect to the current knowledge (dotted lines).  
Bottom: The same for $e^+ e^-$ colliders including a run around the $t\bar{t}$ threshold.}
\label{fig:ew:ST}
\end{figure}

The formalism of the oblique parameters, widely used for the interpretation of EWPO from LEP/SLC, relies on a low-energy expansion of possible new physics effects in the gauge boson self-energies $\Pi_{VV^\prime}(q^2)$:
\begin{equation}
\Pi_{VV^\prime}(q^2)=\Pi_{VV^\prime}(0)+\Pi^\prime_{VV^\prime}(0)q^2 + \frac 12 \Pi^{\prime\prime}_{VV^\prime}(0) q^4 + \dots
\end{equation}
The EWPOs can be approximately described in terms of three such quantities, typically described by the $STU$ parameters, and given by the following combinations of the self-energies and first derivatives evaluated at $q^2=0$:
\begin{equation}
\label{eq:Oblique}
\begin{split}
	\alpha T \equiv &~\frac{1}{c_w^2 m_Z^2}\left(\Pi^{(\rm new)}_{WW}(0) - c_w^2 \Pi^{(\rm new)}_{ZZ}(0)\right)\ , \\ 
	\frac{\alpha}{4\, s_w^{\, 2}c_w^{\, 2}} S \equiv &~\Pi^{\prime~\! (\rm new)}_{ZZ}(0)-\Pi^{\prime~\! (\rm new)}_{\gamma\gamma}(0)
    +\frac{s_w^{\, 2} - c_w^{\, 2}}{c_w s_w} \Pi^{\prime~\! (\rm new)}_{Z\gamma}(0)\ , \\ 
	\frac{\alpha}{4\, s_w^{\, 2}} U \equiv &~\Pi^{\prime~\! (\rm new)}_{WW}(0) - s_w^2 \Pi^{\prime~\! (\rm new)}_{\gamma\gamma}(0) -c_w^2 \Pi^{\prime~\! (\rm new)}_{ZZ}(0) - 2 s_w c_w \Pi^{\prime~\! (\rm new)}_{Z\gamma}(0)\ ,
\end{split}
\end{equation}
where $\Pi^{(\rm new)}$ denotes the contribution from new physics only, and we have denoted by $s_w$ and $c_w$ the sine and cosine of the weak mixing angle, respectively. 
For many new physics scenarios, however, the $U$ parameter is expected to be parametrically suppressed with respect to $S$ and $T$ by a factor $m_W^2/\Lambda^2$, with $\Lambda$ the typical scale of new physics. This is a consequence of $U$ being associated to a dimension-eight operators, whereas $S$ and $T$ can be generated at dimension six. Working at the dimension-six level, the extra degrees of freedom entering in EWPOs can be parameterized by a combination of two additional parameters, called $W$ and $Y$~\cite{Barbieri:2004qk}. These, however, also induce effects that grow with the energy in 2-to-2 fermion processes and are better constrained at high energies, where we currently have a significant amount of data from LEP2 and the LHC. Moreover, in several scenarios, like the simplest composite Higgs models, $W$ and $Y$ are parametrically suppressed with respect to $S$ and $T$. For these reasons, and since a more complete analysis is presented in the SMEFT section, here we restrict ourselves to a simple analysis of EWPOs in terms only of $S$ and $T$. In Table~\ref{tab:STpars} we present the numerical results for the bounds on the $S$ and $T$ parameters presented in Figure 3.3 in Ref.~\cite{deBlas:2025PhysicsBriefingBook}, computed in the {\em aggressive} theory scenario (THA) discussed above. We also include the corresponding limits obtained in the {\em conservative} (THC) scenario and the ones neglecting all theory uncertainties (No TH), as shown in Figure 3.4 in Ref.~\cite{deBlas:2025PhysicsBriefingBook}. For the LCF and FCC--ee, we also present in the table the results obtained before operating at/above the $t\bar{t}$ threshold. 
All these results were obtained working in LEP scheme for the EW precision observables, using as SM inputs $\left\{ \Delta \alpha_{\rm had}^{(5)}(m_Z), \alpha_s(m_Z), m_Z, m_t, m_H\right\}$ ($\alpha(0)$ and $G_F$ are fixed). We include the parametric uncertainties coming from the experimental determination of such inputs, and the intrinsic theory uncertainties in the SM predictions, due to missing higher-order corrections, according to the values given in Tables~\ref{tab:therrZ} and \ref{tab:therrW} for each theory scenario. 
For the future $e^+e^-$ colliders, the results are shown in Figure~\ref{fig:ew:ST} (similar to the lower panel in Figure 3.3 in \cite{deBlas:2025PhysicsBriefingBook}), where we also show for comparison the consequences of not improving the precision of the theory calculations with respect to the current knowledge. 

\begin{table}[htbp]
	\centering
  \begin{tabular}{c c c cccccc}
\toprule
 & TH scenario & HL-LHC & LCF$_{250}$ & LCF$_{350}$ & CLIC & LEP3 & FCC--ee$_{240}$ & FCC--ee\\
\midrule
\midrule
\crowcolorC \cellcolor{white} &No TH  & 0.079 & 0.0061 & 0.0061 & 0.021 & 0.0015 & 0.0007 & 0.0007 \\
                                &THA  & 0.079 & 0.0062 &  0.0062 &0.021 & 0.0018 & 0.0009 & 0.0009 \\
\crowcolorC \cellcolor{white}\multirow{-3}{*}{$S$} &THC  & 0.079 &0.0074 & 0.0074 & 0.021 & 0.0044 &0.0041 &  0.0041 \\
\hline
\crowcolorC \cellcolor{white} &No TH  & 0.054 & 0.0042 &  0.0034 & 0.015 & 0.0014 & 0.0007 & 0.0007 \\
                              &THA  & 0.054 & 0.0050 & 0.0045 & 0.016& 0.0026 & 0.0023 & 0.0009\\
\crowcolorC \cellcolor{white}\multirow{-3}{*}{$T$} &THC  & 0.054 & 0.0062 & 0.0058 &0.016  &0.0050 & 0.0048 & 0.0041 \\
\midrule
\midrule
\crowcolorC \cellcolor{white} &No TH  & 98$\%$ & 57$\%$ & 70$\%$ & 89$\%$ & 64$\%$ & 58$\%$ & 60$\%$ \\
                              &THA & 98$\%$ & 58$\%$ & 65$\%$ & 88$\%$ & 50$\%$ &  46$\%$ & 75$\%$ \\
\crowcolorC \cellcolor{white}\multirow{-3}{*}{$\rho_{ST}$} &THC  & 98$\%$ & 67$\%$ & 72$\%$ & 88$\%$ & 77$\%$ & 77$\%$ & 89$\%$ \\
\bottomrule
\end{tabular}
 	\caption{68$\%$ probability sensitivity to new physics effects in the form of the $S$ and $T$ parameters from the fit to EWPO at different future collider projects. 
    We also provide the correlation, $\rho_{ST}$, between the two parameters. 
    The results are given for the two theory scenarios discussed in this note, as well as neglecting any theory uncertainties. For collider options with projected runs below and above the $t\bar{t}$ threshold, i.e. the LCF and FCC--ee, we report the results obtained with and without the information about the top-quarm mass.
    \label{tab:STpars}}
\end{table}

\subsection{Kappa framework\label{sec:Kappa}}

In the Kappa ($\kappa$) framework~\cite{LHCHiggsCrossSectionWorkingGroup:2012nn,LHCHiggsCrossSectionWorkingGroup:2013rie}, deviations from the SM are parameterized in terms of the so-called coupling-strength modifiers $\kappa_i$. Working in the narrow-width approximation, the rate for any Higgs production and decay process can be written as
\begin{equation}
( \sigma \cdot \mathrm{BR} ) ( i \to \mathrm{H} \to f ) =
\frac{\sigma_i \cdot \Gamma_{H\to f}}{\Gamma_H},
\label{eq:HiggsNWA}
\end{equation}
where $\sigma_i$ is the Higgs production cross section via the initial state $i$, $\Gamma_{H\to f}$ is the partial decay width into the final state $f$, and $\Gamma_H$ is the total Higgs width. The coupling modifiers are defined through
\begin{equation}
( \sigma \cdot \mathrm{BR} ) ( i \to \mathrm{H} \to f ) =
\frac{\sigma_i^{\rm SM}\kappa_i^2 \cdot
\Gamma_{H\to f}^{\rm SM}\kappa_f^2}
{\Gamma_H^{\rm SM}\kappa_H^2
(1-{\rm BR}_{\rm inv}-{\rm BR}_{\rm unt})^{-1}},
\end{equation}
where
\[
\kappa_H^2 \equiv
\sum_j
\frac{\kappa_j^2\,\Gamma_{H\to j}^{\rm SM}}
{\Gamma_H^{\rm SM}}.
\]
In Eq.~\eqref{eq:HiggsNWA} we allow for Higgs decays into invisible states or other new particles giving rise to final states not targeted in current searches, parameterizing their contributions to the total width through ${\rm BR}_{\rm inv}$ and ${\rm BR}_{\rm unt}$, respectively. While ${\rm BR}_{\rm inv}$ can be constrained through dedicated searches, ${\rm BR}_{\rm unt}$ is, by definition, assumed to be experimentally unconstrained.

Within the approximation of Eq.~\eqref{eq:HiggsNWA}, the total Higgs width can be determined indirectly by combining an inclusive production measurement, such as $\sigma_i$, with measurements of the different decay rates. At lepton colliders this is possible through, for example, the recoil-mass method, whereas no analogous observable exists at $pp$ or $ep$ colliders. Consequently, fits for hadron and electron--proton colliders require additional model-dependent assumptions, such as $\kappa_{W,Z}\leq 1$, which is satisfied in many BSM scenarios.

The results of the $\kappa$ fits are presented in Tables~\ref{tab:resultsHadCollKappa}--\ref{tab:resultsLeptCollKappaComb}. The fits for hadron and electron--proton colliders, shown in Table~\ref{tab:resultsHadCollKappa}, impose the constraint $\kappa_{W,Z}\leq 1$, while the fits including lepton-collider measurements, shown in Tables~\ref{tab:resultsLeptCollKappa} and~\ref{tab:resultsLeptCollKappaComb}, do not require this assumption. In all fits, the coupling modifiers for first-generation fermions are fixed to their SM values, since their contribution is negligible at the projected precision. For the non-$e^+e^-$ collider fits, we additionally fix $\kappa_s=1$. Note also that we assume independent coupling modifiers $\kappa_{g}$, $\kappa_{\gamma}$ and $\kappa_{Z\gamma}$ for the radiative processes of ggF/$H\to gg$, $H\to\gamma\gamma$ and $H\to Z\gamma$, respectively.~\footnote{If one assumes there are no additional particles running in the loops generating these processes, $\kappa_{g,\gamma,Z\gamma}$ could be written in terms of the coupling modifiers of the SM fermions and vector bosons.} 
Tables~\ref{tab:resultsHadCollKappa}--\ref{tab:resultsLeptCollKappaComb} provide the numerical values corresponding to Fig.~3.1 of the Physics Briefing Book~\cite{deBlas:2025PhysicsBriefingBook}. Throughout this section, only projected experimental uncertainties are included. The impact of theoretical uncertainties is discussed in the EFT analysis presented below.

\begin{table}[hbt]
\centering
\begin{tabular}{ c | c | cccc }
\toprule
  &  & \multicolumn{4}{c}{HL-LHC+} \\
 & HL-LHC & LHeC & LHeC+FCC--hh & FCC--hh & FCC--hh~(Cons.) \\ 
\midrule
\midrule
$1\geq \kappa_W^{(68\%)} > $ & 0.985 & 0.996 & 0.996 & 0.993 & 0.991 \\ 
$1\geq \kappa_Z^{(68\%)} > $ & 0.987 & 0.993 & 0.994 & 0.994 & 0.992 \\ 
\midrule
$\kappa_g~(\%)$ & 2.0 & 1.6 & 0.85 & 0.77 & 1.1 \\ 
$\kappa_\gamma~(\%)$ & 1.6 & 1.4 & 0.48 & 0.44 & 0.55 \\ 
$\kappa_{Z\gamma}~\!\!(\%)$ & 6.4 & 6.4 & 0.77 & 0.95 & 1.0 \\ 
$\kappa_c~(\%)$ & --- & 3.7 & 3.6 & --- & --- \\ 
$\kappa_t~(\%)$ & 3.2 & 3.2 & 0.8 & 1.3 & 1.9 \\ 
$\kappa_b~(\%)$ & 2.5 & 1.2 & 0.9 & 1.1 & 1.5 \\ 
$\kappa_\mu~(\%)$ & 2.8 & 2.7 & 0.55 & 0.49 & 0.61 \\ 
$\kappa_\tau~(\%)$ & 1.6 & 1.3 & 1.2 & 0.61 & 0.75 \\ 
$\Gamma_H~\!(\%)$ & 3.6 & 1.9 & 1.5 & 1.5 & 2.0 \\ 
\midrule
Br$_\text{inv}^{(95\%)}~(\%) \leq$ & 1.9 & 1.1 & 0.026 & 0.026 & 0.026 \\ 
\arrayrulecolor{black}\bottomrule
\end{tabular}
\caption{$68\%$ probability sensitivity for the $\kappa$ coupling modifiers for the hadron and electron--proton collider scenarios with the constraint $\kappa_{W,Z}\leq1$ imposed. For the first two parameters, $\kappa_W$ and $\kappa_Z$, the lower bound of the $68\%$ probability interval is quoted. 
For Br$_\text{inv}$ the upper bound at 95$\%$ probability is reported.
The "FCC--hh~(Cons.)'' scenario assumes HL-LHC theory uncertainties, while the "FCC--hh'' scenario assumes these uncertainties are reduced by a factor of two. 
\label{tab:resultsHadCollKappa}
}
\end{table}

\begin{sidewaystable}[thp]
\begin{center}
\begin{tabular}{ c |  ccccccccccc }
\toprule
 & \multicolumn{11}{c}{HL-LHC +} \\
 & \multicolumn{3}{c}{LCF} & \multicolumn{3}{c}{CLIC} & LEP3 & \multicolumn{2}{c}{FCC--ee} & FCC & MuC$_{10}$ \\
 & 250 & 550 & 1000 & 380 & 1500 & 3000 &  & 240 & 365 & & \\ 
\midrule
\midrule
$\kappa_W~(\%)$ & 0.88 & 0.2 & 0.18 & 0.25 & 0.18 & 0.17 & 1.0 & 0.63 & 0.29 & 0.28 & 0.26 \\ 
$\kappa_Z~(\%)$ & 0.21 & 0.17 & 0.17 & 0.18 & 0.17 & 0.17 & 0.26 & 0.12 & 0.1 & 0.1 & 0.22 \\ 
$\kappa_g~(\%)$ & 1.0 & 0.56 & 0.46 & 0.68 & 0.56 & 0.5 & 1.2 & 0.69 & 0.49 & 0.44 & 0.49 \\ 
$\kappa_\gamma~(\%)$ & 1.3 & 1.1 & 1.0 & 1.2 & 1.2 & 1.1 & 1.4 & 1.1 & 1.1 & 0.32 & 0.74 \\ 
$\kappa_{Z\gamma}~\!\!(\%)$ & 5.4 & 5.3 & 5.3 & 4.2 & 3.9 & 3.5 & 5.7 & 4.3 & 3.8 & 0.66 & 4.4 \\ 
$\kappa_c~(\%)$ & 1.5 & 0.81 & 0.62 & 0.86 & 0.7 & 0.65 & 1.8 & 0.92 & 0.67 & 0.66 & 1.8 \\ 
$\kappa_t~(\%)$ & 3.1 & 1.9 & 1.2 & 3.1 & 1.7 & 1.6 & 3.1 & 3.1 & 3.1 & 0.75 & 3.1 \\ 
$\kappa_b~(\%)$ & 0.92 & 0.34 & 0.28 & 0.47 & 0.27 & 0.21 & 1.1 & 0.62 & 0.38 & 0.36 & 0.34 \\ 
$\kappa_\mu~(\%)$ & 2.5 & 2.4 & 2.3 & 2.5 & 2.4 & 2.2 & 2.5 & 2.3 & 2.3 & 0.4 & 1.9 \\ 
$\kappa_\tau~(\%)$ & 0.96 & 0.5 & 0.44 & 0.66 & 0.53 & 0.48 & 1.1 & 0.67 & 0.45 & 0.43 & 0.59 \\ 
$\Gamma_H~\!(\%)$ & 1.9 & 0.82 & 0.75 & 1.0 & 0.76 & 0.7 & 2.2 & 1.3 & 0.77 & 0.74 & 1.1 \\ 
\midrule
Br$_\text{inv}^{(95\%)}~(\%) \leq$ & 0.15 & 0.14 & 0.14 & 0.24 & 0.24 & 0.23 & 0.19 & 0.085 & 0.081 & 0.025 & 0.13 \\ 
\bottomrule
\end{tabular}
\caption{$68\%$ probability sensitivity for the $\kappa$ coupling modifiers for $e^+e^-$ and $\mu^+\mu^-$ colliders scenarios. No constraint $\kappa_{W,Z}$ is imposed. For Br$_\text{inv}$ the upper bound at 95$\%$ probabiligy is quoted. The entry labeled by "FCC'' refers to the scenario of the FCC--ee followed by the FCC--hh. See Table~\ref{tab:resultsLeptCollKappaComb} for other combinations.)
\label{tab:resultsLeptCollKappa}
}
\end{center}
\end{sidewaystable}

\begin{table}[ht]
\centering
\setlength\tabcolsep{4.5pt}
\setlength\tabcolsep{2.5pt}
\begin{tabular}{ c |  ccc }
\toprule
\multirow{2}{*}{ } & \multicolumn{3}{c}{HL-LHC+} \\
 & LCF$_{1000}$+MuC$_{10}$ & LEP3+FCC--hh & LEP3+MuC$_{10}$ \\ 
\midrule
\midrule
$\kappa_W~(\%)$ & 0.14 & 0.89 & 0.19 \\ 
$\kappa_Z~(\%)$ & 0.13 & 0.26 & 0.16 \\ 
$\kappa_g~(\%)$ & 0.33 & 0.88 & 0.43 \\ 
$\kappa_\gamma~(\%)$ & 0.65 & 0.41 & 0.71 \\ 
$\kappa_{Z\gamma}~\!\!(\%)$ & 4.0 & 0.73 & 4.1 \\ 
$\kappa_c~(\%)$ & 0.57 & 1.7 & 1.2 \\ 
$\kappa_t~(\%)$ & 1.2 & 0.84 & 3.0 \\ 
$\kappa_b~(\%)$ & 0.2 & 0.84 & 0.28 \\ 
$\kappa_\mu~(\%)$ & 1.8 & 0.48 & 1.9 \\ 
$\kappa_\tau~(\%)$ & 0.36 & 0.91 & 0.48 \\ 
$\Gamma_H~\!(\%)$ & 0.57 & 1.8 & 0.76 \\ 
\midrule
Br$_\text{inv}^{(95\%)}~(\%)\leq $ & 0.095 & 0.026 & 0.11 \\ 
\arrayrulecolor{black}\bottomrule
\end{tabular}
\caption{Same as Table~\ref{tab:resultsLeptCollKappa}, for additional combinations of collider results involving LCF or LEP3.
\label{tab:resultsLeptCollKappaComb}
}
\end{table}

\subsection{Effective Field Theory Framework\label{sec:EFT}}

For the EFT study presented here we employ the SMEFT, where the Higgs belongs to an $SU(2)_L$ doublet. We truncate the effective Lagrangian at dimension six,
\begin{equation}
	{\cal L}_{\rm SMEFT}^{(6)} = {\cal L}_{\rm SM} +
    \sum_i \frac{C_i}{\Lambda^{2}}{\cal O}_i\ ,
\label{stanmodel:eq:smeft}
\end{equation}
where ${\cal O}_i$ are built from the SM fields and symmetries, the Wilson coefficients $C_i$ encode the information about new physics, and $\Lambda$ is the cut-off of the EFT. 
The notation and conventions follow Appendix~A of Ref.~\cite{deBlas:2025PhysicsBriefingBook}. 
Assuming baryon- and lepton-number conservation~\footnote{The latter is implicitly assumed in Eq.~(\ref{stanmodel:eq:smeft}), as otherwise the dimension-five Weinberg operator generating Majorana neutrino masses must also be included.}, the SMEFT contains 2499 dimension-six operators, corresponding to 59 independent operator structures once flavour is ignored. 
We use the so-called {\em Warsaw} basis of dimension-six operators which, for completeness, we show in Table~\ref{tab:dim6BasisAll}. 
Most of these operators involve fermion fields, can induce flavour-changing processes, and are therefore strongly constrained by flavour measurements. To relax such constraints, we assume the new physics respects an $U(2)^5$ flavour symmetry acting on the first two fermion generations, as well as CP conservation. For the left-handed quark sector, the third-generation direction is identified with the physical top-quark direction. These flavour assumptions are imposed at the cutoff scale $\Lambda=10$~TeV, where the heavy new physics is integrated out and the EFT description breaks down. As discussed in Ref.~\cite{deBlas:2025PhysicsBriefingBook}, this choice is motivated by the projected kinematical reach of future high-energy colliders such as the FCC--hh and the 10~TeV muon collider. Altogether, these assumptions reduce the operator basis to 124 independent Wilson coefficients. 

% Basis of dim 6

\begin{table}[p]
  \begin{center}
  {\small
    \begin{tabular}{cclccl}
      \toprule
      & Operator & Notation & & Operator & Notation \\
      \cmidrule{2-3} \cmidrule{5-6}
      % X^3
      \multirow{2}{*}{$X^3$}
      &
      $\varepsilon_{abc} W^{a\,\nu}_\mu W^{b\,\rho}_\nu W^{c\,\mu}_\rho$ &
      $\mathcal{O}_{W}$ &
      &
      $\varepsilon_{abc} \tilde{W}^{a\,\nu}_\mu W^{b\,\rho}_\nu W^{c\,\mu}_\rho$ &
      $\mathcal{O}_{\tilde{W}}$ \\
      &
      $f_{ABC} G^{A\,\nu}_\mu G^{B\,\rho}_\nu G^{C\,\mu}_\rho$ &
      $\mathcal{O}_{G}$ &
      &
      $f_{ABC} \tilde{G}^{A\,\nu}_\mu G^{B\,\rho}_\nu G^{C\,\mu}_\rho$ &
      $\mathcal{O}_{\tilde{G}}$ \\[1mm]
      \midrule
      \vspace{-5mm} & & & & \\
      % phi^6
      $\phi^6$ &
      $(\phi^{\dagger} \phi)^3$ &
      $\mathcal{O}_{\phi}$ &
      & & \\[1mm]
      \midrule
      \vspace{-5mm} & & & & \\
      % phi^4 D^2
      $\phi^4 D^2$ &
      $(\phi^{\dagger} \phi)\square
      (\phi^{\dagger} \phi)$ &
      $\mathcal{O}_{\phi \square}$ &
      &
      $(\phi^{\dagger}D_\mu \phi)
      ((D^\mu \phi)^{\dagger}\phi)$ &
      $\mathcal{O}_{\phi D}$ \\[1mm]
      \midrule
      \vspace{-5mm} & & & & \\
      % psi^2 phi^3
      \multirow{2}{*}{$\psi^2 \phi^2$} &
      $(\phi^{\dagger} \phi)
      \left(\bar{l}_L \phi e_R\right)$ &
      $\mathcal{O}_{e \phi}$ & & \\
      &
      $(\phi^{\dagger} \phi)
      \left(\bar{q}_L \phi d_R\right)$ &
      $\mathcal{O}_{d \phi }$ &
      & 
      $(\phi^{\dagger} \phi)
      (\bar{q}_L \tilde{\phi} u_R)$ &
      $\mathcal{O}_{u \phi}$ \\[1mm]
      \midrule
      \vspace{-5mm} & & & & \\
       % X^2 \phi^2
      \multirow{4}{*}{$X^2 \phi^2$} &
      $\phi^\dagger \phi B_{\mu\nu} B^{\mu\nu} $ &
      $\mathcal{O}_{\phi B}$ &
      &
      $\phi^\dagger \phi \tilde{B}_{\mu\nu} B^{\mu\nu}$ &
      $\mathcal{O}_{\phi \tilde{B}}$ \\
      &
      $\phi^\dagger \phi W_{\mu\nu}^a W^{a\,\mu\nu} $ &
      $\mathcal{O}_{\phi W}$ &
      &
      $\phi^\dagger \phi \tilde{W}_{\mu\nu}^a W^{a\,\mu\nu}$ &
      $\mathcal{O}_{\phi \tilde{W}}$ \\
      &
      $\phi^\dagger \sigma_a \phi W^a_{\mu\nu} B^{\mu\nu}$ &
      $\mathcal{O}_{\phi WB}$ &
      &
      $\phi^\dagger \sigma_a \phi \tilde{W}^a_{\mu\nu} B^{\mu\nu}$ &
      $\mathcal{O}_{\phi\tilde{W}B}$ \\
      &
      $\phi^\dagger \phi G_{\mu\nu}^A G^{A\,\mu\nu} $ &
      $\mathcal{O}_{\phi G}$ &
      &
      $\phi^\dagger \phi \tilde{G}_{\mu\nu}^A G^{A\,\mu\nu}$ &
      $\mathcal{O}_{\phi\tilde{G}}$ \\[1mm]
      \midrule
      \vspace{-5mm} & & & & \\
      % psi^2 X phi
      \multirow{4}{*}{$\psi^2 X \phi$} &
      $\left(\bar{l}_L \sigma^{\mu\nu} e_R\right)
      \phi B_{\mu\nu}$ &
      $\mathcal{O}_{eB}$ &
      &
      $\left(\bar{l}_L \sigma^{\mu\nu} e_R\right)
      \sigma^a \phi W_{\mu\nu}^a$ &
      $\mathcal{O}_{eW}$ \\
      & $\left(\bar{q}_L \sigma^{\mu\nu} u_R\right)
      \tilde{\phi} B_{\mu\nu}$ &
      $\mathcal{O}_{uB}$ &
      &
      $\left(\bar{q}_L \sigma^{\mu\nu} u_R\right)
      \sigma^a \tilde{\phi} W_{\mu\nu}^a$ &
      $\mathcal{O}_{uW}$ \\
      &
      $\left(\bar{q}_L \sigma^{\mu\nu} d_R\right)
      \phi B_{\mu\nu}$ &
      $\mathcal{O}_{dB}$ &
      &
      $\left(\bar{q}_L \sigma^{\mu\nu} d_R\right)
      \sigma^a \phi W_{\mu\nu}^a$ &
      $\mathcal{O}_{dW}$ \\
      &
      $\left(\bar{q}_L \sigma^{\mu\nu} T_A u_R\right)
      \tilde{\phi} G_{\mu\nu}^A$ &
      $\mathcal{O}_{uG}$ &
      &
      $\left(\bar{q}_L \sigma^{\mu\nu} T_A d_R\right)
      \phi G^A_{\mu\nu}$ &
      $\mathcal{O}_{dG}$ \\[1mm]
      \midrule
      \vspace{-5mm} & & & & \\
      % psi^2 phi^2 D
      \multirow{5}{*}{$\psi^2 \phi^2 D$} &
      $(\phi^{\dagger} i\overset{\leftrightarrow}{D}_\mu \phi)
      \left(\bar{l}_L \gamma^\mu l_L\right)$ &
      $\mathcal{O}_{\phi l}^{(1)}$ &
      &
      $(\phi^{\dagger} i \overset{\leftrightarrow}{D}\!^{\!~a}_\mu \phi)
      \left(\bar{l}_L \gamma^\mu \sigma_a l_L\right)$ &
      $\mathcal{O}_{\phi l}^{(3)}$ \\
      &
      $(\phi^{\dagger} i\overset{\leftrightarrow}{D}_\mu \phi)
      \left(\bar{e}_R \gamma^\mu e_R\right)$ &
      $\mathcal{O}_{\phi e}$ &
      & & \\
      &
      $(\phi^{\dagger} i\overset{\leftrightarrow}{D}_\mu \phi)
      \left(\bar{q}_L \gamma^\mu q_L\right)$ &
      $\mathcal{O}_{\phi q}^{(1)}$ &
      &
      $(\phi^{\dagger} i \overset{\leftrightarrow}{D}\!^{\!~a}_\mu \phi)
      \left(\bar{q}_L \gamma^\mu\sigma_a q_L\right)$ &
      $\mathcal{O}_{\phi q}^{(3)}$ \\
      &
      $(\phi^{\dagger} i\overset{\leftrightarrow}{D}_\mu \phi)
      \left(\bar{u}_R \gamma^\mu u_R\right)$ &
      $\mathcal{O}_{\phi u}$ &
      &
      $(\phi^{\dagger} i\overset{\leftrightarrow}{D}_\mu \phi)
      \left(\bar{d}_R \gamma^\mu d_R\right)$ &
      $\mathcal{O}_{\phi d}$ \\
      &
      $(\tilde{\phi}^{\dagger} iD_\mu \phi)
      \left(\bar{u}_R \gamma^\mu d_R\right)$ &
      $\mathcal{O}_{\phi ud}$ &
      & & \\[1mm]
      \midrule
      \vspace{-5mm} & & & & \\
      % 4F: LLLL
      \multirow{3}{*}
      {$\left(\bar{L}L\right)\left(\bar{L}L\right)$} &
      $\left(\bar{l}_L \gamma_\mu l_L\right)
      \left(\bar{l}_L \gamma^\mu l_L\right)$ &
      $\mathcal{O}_{ll}$ &
      & & \\
      &
      $\left(\bar{q}_L \gamma_\mu q_L\right)
      \left(\bar{q}_L \gamma^\mu q_L\right)$ &
      $\mathcal{O}_{qq}^{(1)}$ &
      &
      $\left(\bar{q}_L \gamma_\mu \sigma_a q_L\right)
      \left(\bar{q}_L \gamma^\mu \sigma_a q_L\right)$ &
      $\mathcal{O}_{qq}^{(3)}$ \\
      &
      $\left(\bar{l}_L\gamma_\mu l_L\right)
      \left(\bar{q}_L \gamma^\mu q_L\right)$ &
      $\mathcal{O}_{lq}^{(1)}$ &
      &
      $\left(\bar{l}_L \gamma_\mu\sigma_a l_L\right)
      \left(\bar{q}_L \gamma^\mu\sigma_a q_L\right)$ &
      $\mathcal{O}_{lq}^{(3)}$ \\[1mm]
      \hline
      \vspace{-4mm} & & & & \\
      % 4F: RRRR
      \multirow{4}{*}
      {$\left(\bar{R}R\right)\left(\bar{R}R\right)$} &
      $\left(\bar{e}_R \gamma_\mu e_R\right)
      \left(\bar{e}_R \gamma^\mu e_R\right)$ &
      $\mathcal{O}_{ee}$ &
      & \\
      &
      $\left(\bar{u}_R \gamma_\mu u_R\right)
      \left(\bar{u}_R \gamma^\mu u_R\right)$ &
      $\mathcal{O}_{uu}$ &
      &
      $\left(\bar{d}_R \gamma_\mu d_R\right)
      \left(\bar{d}_R \gamma^\mu d_R\right)$ &
      $\mathcal{O}_{dd}$ \\
      &
      $\left(\bar{u}_R \gamma_\mu u_R\right)
      \left(\bar{d}_R \gamma^\mu d_R\right)$ &
      $\mathcal{O}_{ud}^{(1)}$ &
      &
      $\left(\bar{u}_R \gamma_\mu T_A u_R\right)
      \left(\bar{d}_R \gamma^\mu T_A d_R\right)$ &
      $\mathcal{O}_{ud}^{(8)}$ \\
      &
      $\left(\bar{e}_R \gamma_\mu e_R\right)
      \left(\bar{u}_R \gamma^\mu u_R\right)$ &
      $\mathcal{O}_{eu}$ &
      &
      $\left(\bar{e}_R \gamma_\mu e_R\right)
      \left(\bar{d}_R \gamma^\mu d_R\right)$ &
      $\mathcal{O}_{ed}$ \\[1mm]
      \hline
      \vspace{-4mm} & & & & \\
      % 4F: LLRR
      \multirow{4}{*}
      {$\left(\bar{L}L\right)\left(\bar{R}R\right)$} &
      $\left(\bar{l}_L \gamma_\mu l_L\right)
      \left(\bar{e}_R \gamma^\mu e_R\right)$ &
      $\mathcal{O}_{le}$ &
      &
      $\left(\bar{q}_L \gamma_\mu q_L\right)
      \left(\bar{e}_R \gamma^\mu e_R\right)$
      &
      $\mathcal{O}_{qe}$ \\
      &
      $\left(\bar{l}_L \gamma_\mu l_L\right)
      \left(\bar{u}_R \gamma^\mu u_R\right)$ &
      $\mathcal{O}_{lu}$ &
      &
      $\left(\bar{l}_L \gamma_\mu l_L\right)
      \left(\bar{d}_R \gamma^\mu d_R\right)$ &
      $\mathcal{O}_{ld}$ \\
      &
      $\left(\bar{q}_L \gamma_\mu q_L\right)
      \left(\bar{u}_R \gamma^\mu u_R\right)$ &
      $\mathcal{O}_{qu}^{(1)}$ &
      &
      $\left(\bar{q}_L \gamma_\mu T_A q_L\right)
      \left(\bar{u}_R \gamma^\mu T_A u_R\right)$ &
      $\mathcal{O}_{qu}^{(8)}$ \\
      &
      $\left(\bar{q}_L \gamma_\mu q_L\right)
      \left(\bar{d}_R \gamma^\mu d_R\right)$ &
      $\mathcal{O}_{qd}^{(1)}$ &
      &
      $\left(\bar{q}_L \gamma_\mu T_A q_L\right)
      \left(\bar{d}_R \gamma^\mu T_A d_R\right)$ &
      $\mathcal{O}_{qd}^{(8)}$ \\[1mm]
      \hline
      \vspace{-4mm} & & & & \\
      % 4F:LRRL &  LRLR
      $\left(\bar{L}R\right)\left(\bar{R}L\right)$ &
      $\left(\bar{l}_L e_R\right)
      \left(\bar{d}_R q_L\right)$ &
      $\mathcal{O}_{ledq}$ &
      & & \\[1mm]
      \hline
      \vspace{-4mm} & & & & \\
      \multirow{2}{*}
      {$\left(\bar{L}R\right)\left(\bar{L}R\right)$} &
      $\left(\bar{q}_L  u_R\right) i\sigma_2
      \left(\bar{q}_L d_R\right)^{\mathrm{T}}$ &
      $\mathcal{O}_{quqd}^{(1)}$ &
      &
      $\left(\bar{q}_L T_A u_R\right) i\sigma_2
      \left(\bar{q}_L T_A d_R\right)^{\mathrm{T}}$ &
      $\mathcal{O}_{quqd}^{(8)}$ \\
      &
      $\left(\bar{l}_L  e_R\right) i\sigma_2
      \left(\bar{q}_L u_R\right)^{\mathrm{T}}$ &
      $\mathcal{O}^{(1)}_{lequ}$ &
      &
      $\left(\bar{l}_L  \sigma_{\mu\nu} e_R\right) i\sigma_2
      \left(\bar{q}_L \sigma^{\mu\nu} u_R\right)^{\mathrm{T}}$ &
      $\mathcal{O}^{(3)}_{lequ}$ \\[1mm]
      \bottomrule
    \end{tabular}
    }
    \caption{The Warsaw basis of dimension-six operators~\cite{Grzadkowski:2010es}. Flavour indices are omitted. \label{tab:dim6BasisAll}}
  \end{center}
\end{table}

The predictions for the observables in the fit include all tree-level SMEFT contributions together with the leading-logarithmic corrections obtained from the one-loop renormalization group equations (RGEs) of Refs.~\cite{Jenkins:2013zja,Jenkins:2013wua,Alonso:2013hga}. For several observables that will be measured with high precision at future $e^+e^-$ colliders, namely the EWPO and the $e^+e^- \to ZH$ production cross section, we additionally include the finite NLO contributions from Ref.~\cite{Asteriadis:2024xts}. For other Higgs observables, such as the Higgs decay widths, the finite contributions from the Higgs self-coupling are also included. The complete set of finite NLO corrections for these observables appeared only after the publication of the Physics Briefing Book, see Ref.~\cite{Bellafronte:2026mhp}, and is therefore not included in the results presented there. A consistent inclusion of the remaining finite NLO corrections would also require the resummation of next-to-leading logarithms using the two-loop SMEFT RGEs. These were not available when the Physics Briefing Book~\cite{deBlas:2025PhysicsBriefingBook} was prepared, with the first complete result appearing only recently in Ref.~\cite{Born:2026xkr}.

In practice, several of the 124 operators contribute to the fit only through renormalization-group mixing, and only proportionally to the weak gauge couplings or small Yukawa interactions, resulting in numerically small effects that we neglect. Retaining only operators that enter in precision observables through mixing proportional to the top-quark Yukawa or the strong coupling reduces the fit to 84 Wilson coefficients. The operators excluded from the fit include, in particular, the bottom-quark dipole operators ($({\cal O}_{dV})_{33}$, $V=B,W,G$), which are already strongly constrained by flavour measurements under the flavour assumptions adopted here.

In the calculation of the SMEFT predictions we retain only terms linear in $1/\Lambda^2$, corresponding to the interference between the SM and dimension-six amplitudes, thereby ensuring a consistent treatment within the dimension-six EFT expansion. The calculations are performed in the $G_F$ scheme, changing $\alpha$ from input to predicted observable and using $\{G_F,m_W,m_Z\}$ as electroweak input parameters, in line with most LHC SMEFT analyses and in contrast to the input scheme adopted for the $ST$ analysis in Section~\ref{sec:STpars}.
As in the study of the $ST$ parameters, we also take into account the effect of parametric uncertainties, and include the SM intrinsic uncertainties for the aggressive and conservative theory scenarios following Tables~\ref{tab:therrZ} and \ref{tab:therrW} for the EW observables, Table~\ref{tab:therrH} for the Higgs ones, and 
Table~\ref{tab:therrF1} for difermion production above the $Z$ resonance at $e^+ e^-$ colliders. 

The results presented in the Physics Briefing Book~\cite{deBlas:2025PhysicsBriefingBook} were provided in different forms. First, to provide a comparison in terms of quantities closely related to the experimental observables measured in Higgs and EW observables, the results of the posterior distribution for the Wilson coefficients from the full SMEFT fit was used to obtain the posterior predictions of a series of {\it effective couplings}, defined in terms of physical pseudo-observables. In particular, these are taken as the $Z$- and Higgs-boson partial decay widths and the $Z$-boson asymmetry parameters for each fermion, $A_f$. We define:
\begin{equation}
\Gamma_{Z\to f\bar{f}}=\frac{\alpha~\!m_Z}{6 s_w^2 c_w^2} (|g_{L}^{f}|^2+|g_{R}^{f}|^2),~~~~A_{f}=\frac{|g_{L}^{f}|^2-|g_{R}^{f}|^2}{|g_{L}^{f}|^2+|g_{R}^{f}|^2},
\label{eq:EWEffC}
\end{equation}
\begin{equation}
(g_{HX}^{{\rm eff}})^2=\frac{\Gamma_{H\to X}^{\rm SMEFT}}{\Gamma_{H\to X}^{\rm SM}},
\label{eq:HiggsEffC}
\end{equation}
for the EW and Higgs effective couplings, respectively.~\footnote{In particular, this provides a comparison that is independent of the basis of dimension-six operators chosen for the specific SMEFT analysis, contrary to using the Wilson coefficients or the associated new physics scale.} 
Note that a similar unique definition for the effective Higgs--top-quark coupling in terms of on-shell Higgs observables that can be compared accross different types of colliders is not straightforward. 
Also, to illustrate the sensitivity to multi-boson interactions, we include in our results the precision on the 3 anomalous triple-gauge-couplings $\delta g_{1Z}$, $\delta \kappa_{\gamma}$, $\lambda_{Z}$~\cite{Hagiwara:1986vm}, as well as the Higgs-boson self-interaction, which we denote as $\lambda_3$. Note however that these are just Lagrangian parameters and not observables by themselves. %
Finally, to give a sense of the sensitivity reach to the scales where new particles generating the new effects could be present, we also provide results in terms of the {\it new physics interaction scale} for several operators, defined by $\Lambda/\sqrt{C_i}$.

\begin{figure}[t!]
%\hspace{1.cm}
\includegraphics[width=0.9975\textwidth]{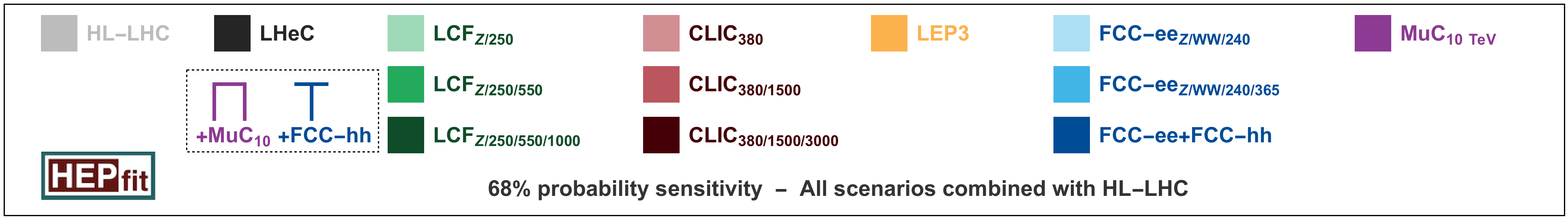}\\
\includegraphics[width=\textwidth]{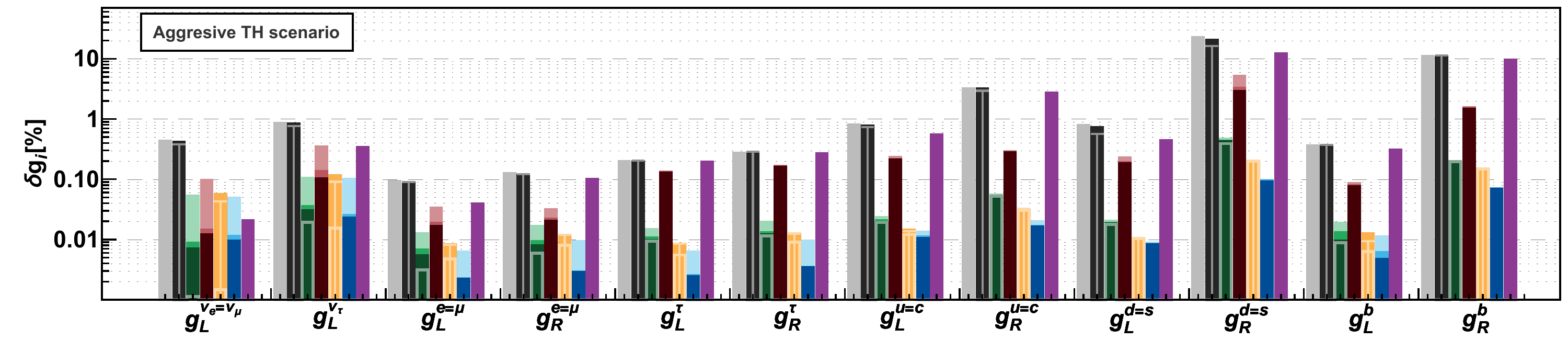}\\
\includegraphics[width=\textwidth]{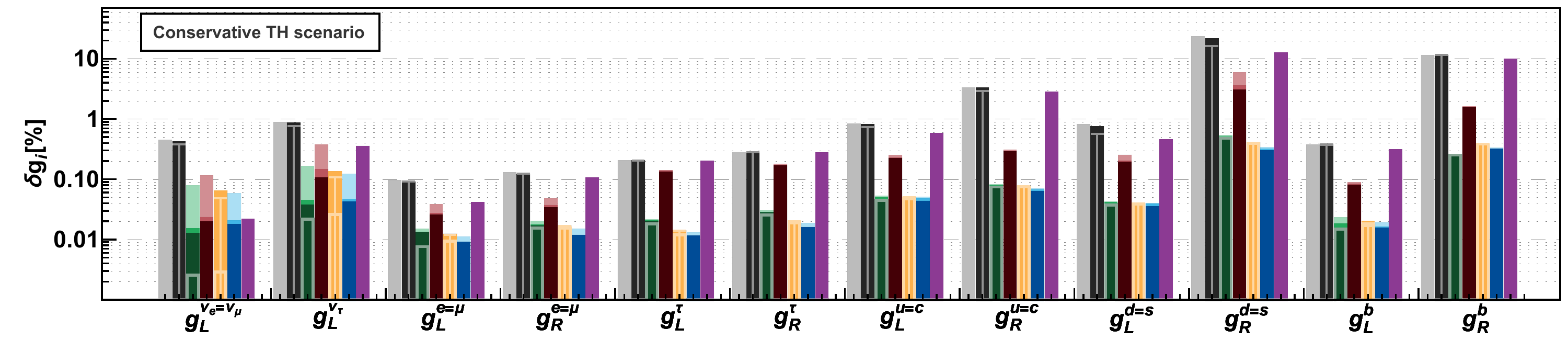}
\caption{(Top panel) 68$\%$ probability sensitivity to the combinations of operators modifying the EW couplings in the dimension-six $U(2)^5$-symmetric SMEFT framework in the {\it aggressive} theory scenario, as in Figure 3.5 of Ref.~\cite{deBlas:2025PhysicsBriefingBook}. (Bottom panel) The same for the {\it conservative} theory scenario. 
The numerical values can be read from Table~\ref{tab:eft-global-EWgi1}.
}
\label{fig:eft:gZffeff_THA_THC}
\end{figure}

The numerical results for the couplings shown in Figures 3.5, 3.6 and 3.7 of Ref.~\cite{deBlas:2025PhysicsBriefingBook} are given in Tables~\ref{tab:eft-global-Higgsgi1}-\ref{tab:eft-global-EWgi1Comb}, both for the aggressive and conservative theory scenarios. 
The results in both scenarios are shown in Figure~\ref{fig:eft:gZffeff_THA_THC} for the effective EW couplings, and in Figure~\ref{eq:HiggsEffC} for the effective Higgs couplings and aTGC. (For the Higgs cubic interaction, Figure 3.7 of~\cite{deBlas:2025PhysicsBriefingBook} already compares the results with different assumptions for the theory uncertainties.)
Similarly, the reach on the interaction scale for the bosonic and two-fermion operators modifying the EWPO at tree level are given in Tables~\ref{tab:eft-global-OiEW1}-\ref{tab:eft-global-OiEW1Comb}~\footnote{We note that, although not shown here, in the chosen scheme for the SM input parameters the four-left-handed-leptons operator also contribute indirectly to all EW observables through its modifications to muon decay. This operator can also be directly tested in $2\to 2$ lepton processes, where it benefits from operation at high centre-of-mass energies. The bounds from $2\to 2$ fermion processes off-the-pole are discussed separately.}, and the ones for the operators modifying the Top-quark processes at lepton colliders, are given in Tables~\ref{tab:eft-global-OiTop1}-\ref{tab:eft-global-OiTop1Comb}. In this last table, the results are presented in terms of the conventions used by the {\em LHC Top Working Group} in~\cite{Aguilar-Saavedra:2018ksv}, related to that in Table~\ref{tab:dim6BasisAll} by
\begin{eqnarray}
    C_{Q l}^{\pm}&=&(C_{l q}^{(1)})_{33} \pm (C_{l q}^{(3)})_{33},\nonumber\\
    C_{Q l}^{(3)}&=&(C_{l q}^{(3)})_{33},\nonumber\\
    C_{t e}&=&(C_{e u})_{\ell\ell 33},\nonumber\\
    C_{t l}&=&(C_{l u})_{\ell\ell 33},\nonumber\\
    C_{Q e}&=&(C_{q e})_{3 3 \ell\ell},\nonumber\\
    C_{\phi Q}^{\pm}&=&(C_{\phi q}^{(1)})_{33} \pm (C_{\phi q}^{(3)})_{33},\nonumber\\
    C_{\phi Q}^{(3)}&=&(C_{\phi q}^{(3)})_{33},\label{eq:SMEFTLHCTopWG}\\    
    C_{\phi t}&=&(C_{\phi u})_{33},\nonumber\\
    C_{t \phi}&=&(C_{u \phi})_{33},\nonumber\\
    C_{tZ}&=&-s_w (C_{u B})_{33} + c_w (C_{u W})_{33},\nonumber\\
    C_{tW}&=&(C_{uW})_{33}.\nonumber
\end{eqnarray}
In Tables~\ref{tab:eft-global-OiEW1}-\ref{tab:eft-global-OiTop1Comb}, for the limit on the interaction scale of the different operators, we present both the results from the global fit and, in parentheses, the ones from fits where only a single operator is assumed to be present at the scale $\Lambda$. 
For the aggressive theory scenario, these correspond to the bottom panel in Figure 3.5 and to Figure 3.8 in Ref.~\cite{deBlas:2025PhysicsBriefingBook}, for the EW and Top-quark operators, respectively. For completeness, we also provide the results for the conservative theory scenario.

\begin{figure}[t!]
%\hspace{1.cm}
\includegraphics[width=0.9975\textwidth]{Figures/Leg_General_noLogo.pdf}\\
\includegraphics[width=\textwidth]{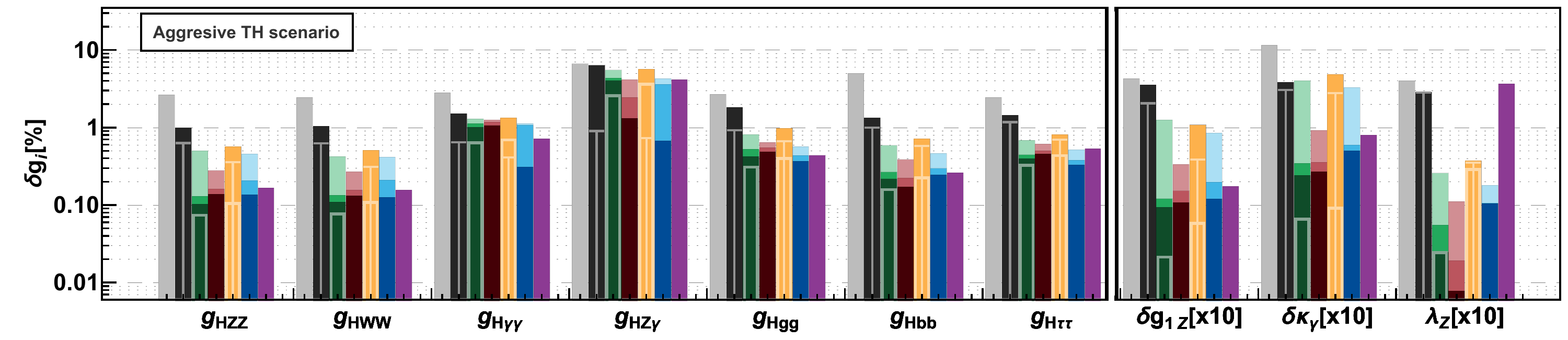}\\
\includegraphics[width=\textwidth]{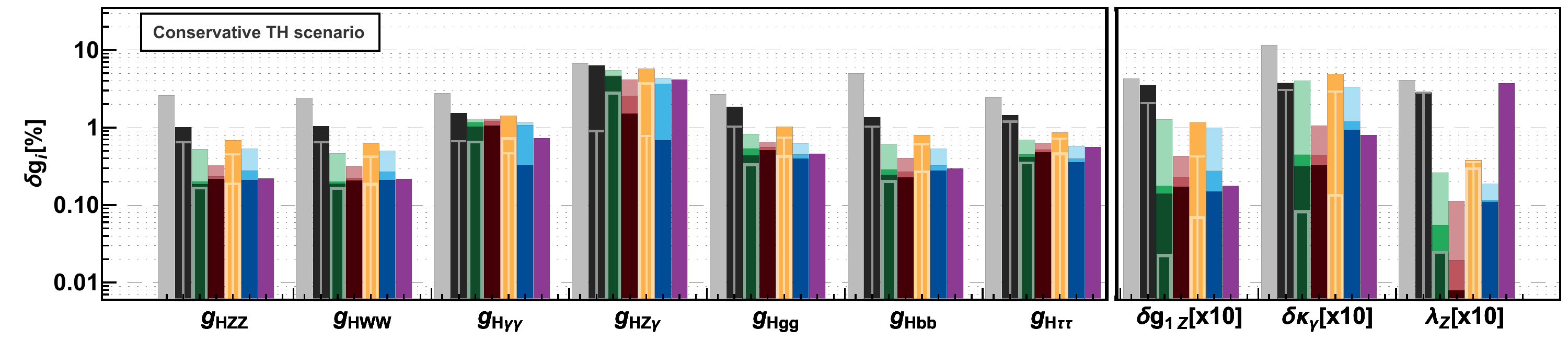}
\caption{(Top panel) 68$\%$ probability sensitivity to the effective Higgs couplings and aTGCs in the dimension-six $U(2)^5$-symmetric SMEFT framework in the {\it aggressive} theory scenario, as in Figure 3.6 of Ref.~\cite{deBlas:2025PhysicsBriefingBook}. (Bottom panel) The same for the {\it conservative} theory scenario. 
The numerical values can be read from Table~\ref{tab:eft-global-Higgsgi1}.
}
\label{fig:eft:gHeff_THA_THC}
\end{figure}

Finally, in order to illustrate the interplay between high-energy measurements and precision constraints on new effective interactions, in Figure 3.9 of Ref.~\cite{deBlas:2025PhysicsBriefingBook} we presented the reach on the interaction scale for several operators contributing to di-boson or di-fermion processes at lepton and hadron colliders. The definition of these operators can be read from the Lagrangaian in Eq.~(A.4) of Appendix A in Ref.~\cite{deBlas:2025PhysicsBriefingBook}. This Lagrangian describes, in particular, the low-energy limit of a general class of composite Higgs models under the assumptions in Refs.~\cite{Giudice:2007fh,Liu:2016idz}. For completeness, we present the corresponding Lagrangian also here:
\begin{equation}
\begin{split}
{\cal L}_{\mathrm{SILH}}=&
\frac{c_\phi }{\Lambda^2} \partial_\mu (\phi^\dagger \phi)\partial^\mu (\phi^\dagger \phi)+ 
\frac{c_T}{\Lambda^2} (\phi^\dagger \overset{\leftrightarrow}{D}_\mu \phi)(\phi^\dagger \overset{\leftrightarrow}{D}{~\!}^\mu \phi) - 
\frac{c_6}{\Lambda^2} (\phi^\dagger \phi)^3+\\
&
+\left(\frac{c_{y_f} }{\Lambda^2} y^f_{ij} \phi^\dagger \phi  \bar{\psi}_{Li} \phi \psi_{Rj} + {\rm h.c.} \right)\\
&
+\frac{c_W}{\Lambda^2} \left(\phi^\dagger i \overset{\leftrightarrow}{D}\!^{\!~a}_\mu \phi\right) D_\nu W^{a~\!\mu\nu} +\frac{c_B}{\Lambda^2} \left(\phi^\dagger i \overset{\leftrightarrow}{D}_\mu \phi\right) \partial_\nu B^{\mu\nu}\\
&
+\frac{c_{\phi W}}{\Lambda^2} i D_\mu\phi^\dagger \sigma_a D_\nu \phi W^{a~\!\mu\nu}+\frac{c_{\phi B} }{\Lambda^2} i D_\mu\phi^\dagger D_\nu \phi B^{\mu\nu}\\
&+\frac{c_{\gamma} }{\Lambda^2}  \phi^\dagger \phi B^{\mu\nu}B_{\mu\nu}+\frac{c_{g}}{\Lambda^2}  \phi^\dagger \phi G^{A~\!\mu\nu}G^A_{\mu\nu}\\
&-\frac{c_{2W}}{\Lambda^2} (D^\mu W_{\mu\nu}^a)(D_\rho W^{a~\!\rho\nu})-\frac{c_{2B}}{\Lambda^2} (\partial^\mu B_{\mu\nu})(\partial_\rho B^{\rho\nu})-\frac{c_{2G}}{\Lambda^2} (D^\mu G_{\mu\nu}^A)(D_\rho G^{A~\!\rho\nu})\\
&+\frac{c_{3W}}{ \Lambda^2} \varepsilon_{abc}W_{\mu}^{a~\nu}W_\nu^{b~\rho}W_\rho^{c~\mu}+\frac{c_{3G}}{ \Lambda^2}  f_{ABC}G_{\mu}^{A~\nu}G_\nu^{B~\rho}G_\rho^{C~\mu}.
\end{split}
\label{eq:LSilh}
\end{equation}
The numerical bounds in Figure 3.9 of Ref.~\cite{deBlas:2025PhysicsBriefingBook} are presented in Tables~\ref{tab:eft-global-OiSILH}-\ref{tab:eft-global-OiSILHComb}, which also includes the numbers for the conservative theory scenario, not shown in the previous reference. As in the tables above, we give the values of the single-operator limits in parenthesis.

\section*{Acknowledgments}
The work of J.B. has been partially funded by MICIU/AEI/\allowbreak10.13039/501100011033 and 
FEDER/UE (grants PID2022-139466NB-C21 and PID2025-170216NB-I00). 
P.P.G. is supported by the Ramón y Cajal grant~RYC2022-038517-I funded by MCIN/AEI/10.13039/501100011033 and by ESF+ by the grant PID2025-172338NB-I00 funded by MICIU/AEI/
10.13039/501100011033 and by ERDF/EU,  and by the R\&D\&I Project CEX2025-001574-S, funded by MICIU/AEI/10.13039/\allowbreak501100011033 within the Particle Physics in the Standard Model and Beyond research line.

\clearpage

\section*{Acronyms}
\begin{acronym}
    \acro{FCC}{Future Circular Collider}
    \acro{CLIC}{Compact Linear Collider}
    \acro{LCF}{Linear Collider Facility at CERN}
    \acro{LHC}{Large Hadron Collider}
    \acro{HLLHC}[HL-LHC]{high luminosity LHC}
    \acro{EW}{electroweak}
    \acro{EFT}{effective field theory}
    \acro{SMEFT}{Standard Model effective field theory}
    \acro{BSM}{beyond the Standard Model}
    \acro{SM}{Standard Model}
    \acro{PDF}{Parton Density Function}
    \acro{VBF}{Vector Boson Fusion}
    \acro{EWSB}{Electroweak Symmetry Breaking}
    \acro{QCD}{Quantum Chromodynamics}
    \acro{EWPO}{Electroweak Precision Observables}
    \acro{MC}{Monte-Carlo}
    \acro{PPG EW WG}{Electroweak physics working group of the Physics Preparatory Group}
    \acro{NLO}{next-to-leading order}
    \acro{NNLO}{next-to-next-to-leading order}
\end{acronym}

\clearpage

\begin{sidewaystable}[thp]
\begin{center}
\begin{tabular}{ c | c | c | c | ccc | ccc | c | ccc | c}
\toprule
&   & &\multicolumn{12}{c}{HL-LHC +}  \\
& TH & HL-LHC & LHeC & \multicolumn{3}{c|}{LCF} & \multicolumn{3}{c|}{CLIC} & LEP3 & \multicolumn{2}{c}{FCC--ee} & FCC & MuC$_{10}$ \\
& scenario    &  & & 250 & 550 & 1000 & 380 & 1500 & 3000 &  & 240 & 365 &  &  \\ 
\midrule\midrule
& THA & 2.6 & 1.0 & 0.5 & 0.13 & 0.1 & 0.28 & 0.16 & 0.14 & 0.57 & 0.46 & 0.21 & 0.14 & 0.17 \\ 
\crowcolorC \cellcolor{white}\multirow{-2}{*}{$g_{HZZ}$} & THC & 2.6 & 1.0 & 0.53 & 0.2 & 0.19 & 0.33 & 0.23 & 0.22 & 0.69 & 0.54 & 0.28 & 0.21 & 0.22 \\ 
 \hline 
 & THA & 2.4 & 1.0 & 0.42 & 0.13 & 0.11 & 0.27 & 0.16 & 0.13 & 0.51 & 0.42 & 0.21 & 0.13 & 0.16 \\ 
\crowcolorC \cellcolor{white}\multirow{-2}{*}{$g_{HWW}$} & THC & 2.4 & 1.0 & 0.47 & 0.2 & 0.19 & 0.32 & 0.22 & 0.21 & 0.62 & 0.5 & 0.27 & 0.21 & 0.22 \\ 
 \hline 
 & THA & 2.8 & 1.5 & 1.3 & 1.1 & 1.0 & 1.3 & 1.2 & 1.1 & 1.3 & 1.1 & 1.1 & 0.31 & 0.72 \\ 
\crowcolorC \cellcolor{white}\multirow{-2}{*}{$g_{H\gamma\gamma}$} & THC & 2.8 & 1.5 & 1.3 & 1.2 & 1.0 & 1.3 & 1.2 & 1.1 & 1.4 & 1.2 & 1.1 & 0.33 & 0.73 \\ 
 \hline 
 & THA & 6.7 & 6.4 & 5.5 & 4.4 & 4.1 & 4.2 & 2.5 & 1.3 & 5.7 & 4.3 & 3.6 & 0.67 & 4.1 \\ 
\crowcolorC \cellcolor{white}\multirow{-2}{*}{$g_{HZ\gamma}$} & THC & 6.7 & 6.3 & 5.5 & 4.7 & 4.6 & 4.1 & 2.6 & 1.5 & 5.7 & 4.4 & 3.7 & 0.69 & 4.2 \\ 
 \hline 
 & THA & 2.7 & 1.8 & 0.82 & 0.53 & 0.42 & 0.65 & 0.56 & 0.49 & 0.98 & 0.57 & 0.44 & 0.37 & 0.44 \\ 
\crowcolorC \cellcolor{white}\multirow{-2}{*}{$g_{Hgg}$} & THC & 2.7 & 1.9 & 0.83 & 0.54 & 0.44 & 0.66 & 0.56 & 0.51 & 1.0 & 0.63 & 0.45 & 0.4 & 0.46 \\ 
 \hline 
 & THA & 5.0 & 1.3 & 0.59 & 0.27 & 0.22 & 0.39 & 0.22 & 0.17 & 0.72 & 0.47 & 0.3 & 0.25 & 0.26 \\ 
\crowcolorC \cellcolor{white}\multirow{-2}{*}{$g_{Hbb}$} & THC & 5.0 & 1.4 & 0.61 & 0.29 & 0.25 & 0.41 & 0.27 & 0.23 & 0.8 & 0.53 & 0.32 & 0.28 & 0.3 \\ 
 \hline 
 & THA & 2.4 & 1.4 & 0.69 & 0.45 & 0.4 & 0.61 & 0.5 & 0.46 & 0.81 & 0.52 & 0.38 & 0.33 & 0.54 \\ 
\crowcolorC \cellcolor{white}\multirow{-2}{*}{$g_{H\tau\tau}$} & THC & 2.4 & 1.4 & 0.7 & 0.45 & 0.42 & 0.62 & 0.52 & 0.48 & 0.87 & 0.58 & 0.4 & 0.36 & 0.56 \\ 
 \hline 
 & THA & 27 & 27 & 27 & 9.1 & 6.7 & 27 & 22* & 8.7* & 26 & 26 & 17 & 3.2 & 3.7 \\ 
\crowcolorC \cellcolor{white}\multirow{-2}{*}{$\lambda_{3}$} & THC & 27 & 27 & 27 & 9.4 & 6.8 & 27 & 22* & 8.7* & 26 & 26 & 19 & 3.3 & 3.8 \\ 
 \hline 
 & THA & 0.43 & 0.36 & 0.13 & 0.012 & 0.0094 & 0.033 & 0.015 & 0.011 & 0.11 & 0.086 & 0.02 & 0.012 & 0.018 \\ 
\crowcolorC \cellcolor{white}\multirow{-2}{*}{$\delta g_{1Z}$} & THC & 0.43 & 0.35 & 0.13 & 0.018 & 0.014 & 0.043 & 0.023 & 0.017 & 0.12 & 0.1 & 0.028 & 0.015 & 0.018 \\ 
 \hline 
 & THA & 1.2 & 0.38 & 0.4 & 0.034 & 0.024 & 0.092 & 0.036 & 0.027 & 0.48 & 0.33 & 0.059 & 0.05 & 0.08 \\ 
\crowcolorC \cellcolor{white}\multirow{-2}{*}{$\delta \kappa_{\gamma}$} & THC & 1.2 & 0.38 & 0.4 & 0.045 & 0.031 & 0.11 & 0.043 & 0.033 & 0.49 & 0.33 & 0.12 & 0.093 & 0.08 \\ 
 \hline 
 & THA & 0.41 & 0.29 & 0.026 & 0.0056 & 0.0025 & 0.011 & 0.0019 & 0.0008 & 0.037 & 0.018 & 0.011 & 0.01 & 0.37 \\ 
\crowcolorC \cellcolor{white}\multirow{-2}{*}{$\lambda_{Z}$} & THC & 0.41 & 0.28 & 0.026 & 0.0056 & 0.0024 & 0.011 & 0.0019 & 0.0008 & 0.038 & 0.019 & 0.012 & 0.011 & 0.37 \\ 
\bottomrule
\end{tabular}
\caption{\label{tab:eft-global-Higgsgi1} 
68\% probability sensivity to modifications of the effective Higgs couplings, Higgs selfcoupling and aTGC in the dimension-six $U(2)^5$-symmetric SMEFT framework. Results are given for the {\it aggressive} (THA) and {\it conservative} (THC) theory scenarios. The CLIC results for $\lambda_3$ have been marked with an asterisk (*) to indicate that, unlike the LCF ones, where the di-Higgs inputs were updated in 2025~\cite{Berggren:2025fpw}, these still come from the 2019 study in~\cite{Roloff:2019crr} and may also benefit from a reanalysis with more modern techniques. 
}
\end{center}
\end{sidewaystable}

\begin{table}[ht]
\centering
\setlength\tabcolsep{4.5pt}
\setlength\tabcolsep{2.5pt}
\begin{tabular}{ c | c | cccc }
\toprule
& TH & \multicolumn{4}{c}{HL-LHC+} \\
 & scenario & LHeC+FCC--hh & LCF$_{1000}$+MuC$_{10}$ & LEP3+FCC--hh & LEP3+MuC$_{10}$ \\ 
\midrule
\midrule
 & THA & 0.63 & 0.074 & 0.36 & 0.1 \\ 
\crowcolorC \cellcolor{white}\multirow{-2}{*}{$g_{HZZ}$} & THC & 0.64 & 0.17 & 0.45 & 0.19 \\ 
 \hline 
 & THA & 0.62 & 0.076 & 0.31 & 0.11 \\ 
\crowcolorC \cellcolor{white}\multirow{-2}{*}{$g_{HWW}$} & THC & 0.65 & 0.16 & 0.42 & 0.19 \\ 
 \hline 
 & THA & 0.65 & 0.63 & 0.41 & 0.69 \\ 
\crowcolorC \cellcolor{white}\multirow{-2}{*}{$g_{H\gamma\gamma}$} & THC & 0.67 & 0.65 & 0.47 & 0.72 \\ 
 \hline 
 & THA & 0.9 & 2.6 & 0.73 & 3.6 \\ 
\crowcolorC \cellcolor{white}\multirow{-2}{*}{$g_{HZ\gamma}$} & THC & 0.9 & 2.8 & 0.77 & 3.7 \\ 
 \hline 
 & THA & 0.92 & 0.31 & 0.67 & 0.4 \\ 
\crowcolorC \cellcolor{white}\multirow{-2}{*}{$g_{Hgg}$} & THC & 1.0 & 0.33 & 0.74 & 0.42 \\ 
 \hline 
 & THA & 1.0 & 0.16 & 0.58 & 0.22 \\ 
\crowcolorC \cellcolor{white}\multirow{-2}{*}{$g_{Hbb}$} & THC & 1.0 & 0.2 & 0.61 & 0.27 \\ 
 \hline 
 & THA & 1.2 & 0.33 & 0.69 & 0.43 \\ 
\crowcolorC \cellcolor{white}\multirow{-2}{*}{$g_{H\tau\tau}$} & THC & 1.2 & 0.35 & 0.72 & 0.46 \\ 
 \hline 
 & THA & 5.9 & 2.6 & 4.1 & 3.7 \\ 
\crowcolorC \cellcolor{white}\multirow{-2}{*}{$\lambda_{3}$} & THC & 5.8 & 2.7 & 4.1 & 3.7 \\ 
 \hline 
 & THA & 0.21 & 0.0021 & 0.039 & 0.0058 \\ 
\crowcolorC \cellcolor{white}\multirow{-2}{*}{$\delta g_{1Z}$} & THC & 0.21 & 0.0022 & 0.042 & 0.0069 \\ 
 \hline 
 & THA & 0.31 & 0.0066 & 0.28 & 0.0091 \\ 
\crowcolorC \cellcolor{white}\multirow{-2}{*}{$\delta \kappa_{\gamma}$} & THC & 0.31 & 0.0082 & 0.29 & 0.013 \\ 
 \hline 
 & THA & 0.28 & 0.0024 & 0.035 & 0.029 \\ 
\crowcolorC \cellcolor{white}\multirow{-2}{*}{$\lambda_{Z}$} & THC & 0.28 & 0.0025 & 0.036 & 0.029 \\ 

\arrayrulecolor{black}\bottomrule
\end{tabular}
\caption{Same as Table~\ref{tab:eft-global-Higgsgi1}, for additional combinations of collider results involving LHeC, LCF or LEP3.
\label{tab:eft-global-Higgsgi1Comb}
}
\end{table}

\begin{sidewaystable}[thp]
\begin{center}
\begin{tabular}{ c | c | c | c | ccc | ccc | c | ccc | c}
\toprule
&    & &\multicolumn{12}{c}{HL-LHC +}  \\
& TH & HL-LHC & LHeC & \multicolumn{3}{c|}{LCF} & \multicolumn{3}{c|}{CLIC} & LEP3 & \multicolumn{2}{c}{FCC--ee} & FCC & MuC$_{10}$ \\
& scenario    &  &  & 250 & 550 & 1000 & 380 & 1500 & 3000 &  & 240 & 365 &  &  \\ 
\midrule\midrule
 & THA & 0.45 & 0.43 & 0.056 & 0.0093 & 0.0075 & 0.1 & 0.015 & 0.013 & 0.06 & 0.051 & 0.012 & 0.01 & 0.022 \\ 
\crowcolorC \cellcolor{white}\multirow{-2}{*}{$g_{L}^{\nu_e=\nu_\mu}$} & THC & 0.45 & 0.43 & 0.079 & 0.015 & 0.013 & 0.12 & 0.024 & 0.02 & 0.065 & 0.059 & 0.021 & 0.018 & 0.022 \\ 
 \hline 
 & THA & 0.9 & 0.87 & 0.11 & 0.037 & 0.032 & 0.36 & 0.14 & 0.11 & 0.12 & 0.11 & 0.027 & 0.024 & 0.36 \\ 
\crowcolorC \cellcolor{white}\multirow{-2}{*}{$g_{L}^{\nu_\tau}$} & THC & 0.9 & 0.87 & 0.17 & 0.046 & 0.038 & 0.38 & 0.15 & 0.11 & 0.14 & 0.12 & 0.047 & 0.043 & 0.36 \\ 
 \hline 
 & THA & 0.097 & 0.093 & 0.013 & 0.0071 & 0.0057 & 0.035 & 0.02 & 0.017 & 0.0086 & 0.0065 & 0.0024 & 0.0023 & 0.041 \\ 
\crowcolorC \cellcolor{white}\multirow{-2}{*}{$g_{L}^{e=\mu}$} & THC & 0.098 & 0.092 & 0.015 & 0.013 & 0.013 & 0.039 & 0.028 & 0.026 & 0.013 & 0.011 & 0.0091 & 0.0091 & 0.042 \\ 
 \hline 
 & THA & 0.13 & 0.12 & 0.017 & 0.0098 & 0.0083 & 0.033 & 0.023 & 0.021 & 0.012 & 0.0098 & 0.0031 & 0.003 & 0.11 \\ 
\crowcolorC \cellcolor{white}\multirow{-2}{*}{$g_{R}^{e=\mu}$} & THC & 0.13 & 0.12 & 0.021 & 0.018 & 0.017 & 0.048 & 0.038 & 0.034 & 0.017 & 0.015 & 0.012 & 0.012 & 0.11 \\ 
 \hline 
 & THA & 0.21 & 0.21 & 0.016 & 0.011 & 0.01 & 0.14 & 0.14 & 0.13 & 0.0088 & 0.0065 & 0.0027 & 0.0026 & 0.2 \\ 
\crowcolorC \cellcolor{white}\multirow{-2}{*}{$g_{L}^{\tau}$} & THC & 0.21 & 0.21 & 0.022 & 0.021 & 0.02 & 0.14 & 0.14 & 0.14 & 0.014 & 0.013 & 0.012 & 0.012 & 0.2 \\ 
 \hline 
 & THA & 0.28 & 0.28 & 0.02 & 0.014 & 0.013 & 0.17 & 0.17 & 0.17 & 0.013 & 0.0099 & 0.0037 & 0.0036 & 0.28 \\ 
\crowcolorC \cellcolor{white}\multirow{-2}{*}{$g_{R}^{\tau}$} & THC & 0.28 & 0.28 & 0.03 & 0.029 & 0.028 & 0.18 & 0.17 & 0.17 & 0.02 & 0.019 & 0.016 & 0.016 & 0.28 \\ 
 \hline 
 & THA & 0.84 & 0.82 & 0.024 & 0.022 & 0.02 & 0.25 & 0.23 & 0.22 & 0.015 & 0.014 & 0.012 & 0.011 & 0.58 \\ 
\crowcolorC \cellcolor{white}\multirow{-2}{*}{$g_{L}^{u=c}$} & THC & 0.84 & 0.82 & 0.054 & 0.051 & 0.047 & 0.25 & 0.23 & 0.22 & 0.051 & 0.051 & 0.048 & 0.044 & 0.58 \\ 
 \hline 
 & THA & 3.3 & 3.3 & 0.058 & 0.055 & 0.054 & 0.31 & 0.29 & 0.29 & 0.033 & 0.021 & 0.017 & 0.017 & 2.8 \\ 
\crowcolorC \cellcolor{white}\multirow{-2}{*}{$g_{R}^{u=c}$} & THC & 3.4 & 3.3 & 0.082 & 0.08 & 0.076 & 0.31 & 0.3 & 0.29 & 0.077 & 0.071 & 0.068 & 0.064 & 2.8 \\ 
 \hline 
 & THA & 0.83 & 0.76 & 0.021 & 0.02 & 0.019 & 0.24 & 0.2 & 0.19 & 0.011 & 0.0092 & 0.0089 & 0.0086 & 0.46 \\ 
\crowcolorC \cellcolor{white}\multirow{-2}{*}{$g_{L}^{d=s}$} & THC & 0.83 & 0.76 & 0.043 & 0.042 & 0.04 & 0.26 & 0.2 & 0.19 & 0.04 & 0.04 & 0.039 & 0.036 & 0.46 \\ 
 \hline 
 & THA & 24 & 21 & 0.5 & 0.46 & 0.44 & 5.5 & 3.4 & 3.0 & 0.2 & 0.1 & 0.1 & 0.096 & 13 \\ 
\crowcolorC \cellcolor{white}\multirow{-2}{*}{$g_{R}^{d=s}$} & THC & 24 & 22 & 0.54 & 0.51 & 0.51 & 6.0 & 3.6 & 3.1 & 0.41 & 0.34 & 0.32 & 0.3 & 13 \\ 
 \hline 
 & THA & 0.38 & 0.37 & 0.02 & 0.014 & 0.01 & 0.089 & 0.084 & 0.08 & 0.013 & 0.012 & 0.0064 & 0.005 & 0.32 \\ 
\crowcolorC \cellcolor{white}\multirow{-2}{*}{$g_{L}^{b}$} & THC & 0.38 & 0.38 & 0.023 & 0.018 & 0.016 & 0.089 & 0.085 & 0.081 & 0.02 & 0.019 & 0.016 & 0.016 & 0.32 \\ 
 \hline 
 & THA & 12 & 11 & 0.2 & 0.2 & 0.2 & 1.6 & 1.6 & 1.5 & 0.15 & 0.074 & 0.073 & 0.073 & 10 \\ 
\crowcolorC \cellcolor{white}\multirow{-2}{*}{$g_{R}^{b}$} & THC & 12 & 12 & 0.26 & 0.26 & 0.26 & 1.6 & 1.6 & 1.6 & 0.39 & 0.33 & 0.32 & 0.32 & 10 \\ 
 \hline 
\bottomrule
\end{tabular}
\caption{\label{tab:eft-global-EWgi1} 
68\% probability sensivity to modifications of the effective fermionic electroweak couplings in the dimension-six $U(2)^5$-symmetric SMEFT framework. Results are given for the {\it aggressive} (THA) and {\it conservative} (THC) theory scenarios. 
}
\end{center}
\end{sidewaystable}

\begin{table}[ht]
\centering
\setlength\tabcolsep{4.5pt}
\setlength\tabcolsep{2.5pt}
\begin{tabular}{ c | c | cccc }
\toprule
& TH & \multicolumn{4}{c}{HL-LHC+} \\
 & scenario & LHeC+FCC--hh & LCF$_{1000}$+MuC$_{10}$ & LEP3+FCC--hh & LEP3+MuC$_{10}$ \\ 
\midrule
\midrule
 & THA & 0.38 & 0.0011 & 0.043 & 0.0015 \\ 
\crowcolorC \cellcolor{white}\multirow{-2}{*}{$g_{L}^{\nu_e=\nu_\mu}$} & THC & 0.38 & 0.0026 & 0.049 & 0.0029 \\ 
 \hline 
 & THA & 0.77 & 0.019 & 0.091 & 0.015 \\ 
\crowcolorC \cellcolor{white}\multirow{-2}{*}{$g_{L}^{\nu_\tau}$} & THC & 0.77 & 0.022 & 0.11 & 0.026 \\ 
 \hline 
 & THA & 0.09 & 0.0031 & 0.0081 & 0.0047 \\ 
\crowcolorC \cellcolor{white}\multirow{-2}{*}{$g_{L}^{e=\mu}$} & THC & 0.091 & 0.0076 & 0.012 & 0.0094 \\ 
 \hline 
 & THA & 0.12 & 0.0059 & 0.012 & 0.008 \\ 
\crowcolorC \cellcolor{white}\multirow{-2}{*}{$g_{R}^{e=\mu}$} & THC & 0.12 & 0.015 & 0.017 & 0.015 \\ 
 \hline 
 & THA & 0.20 & 0.0094 & 0.0085 & 0.0055 \\ 
\crowcolorC \cellcolor{white}\multirow{-2}{*}{$g_{L}^{\tau}$} & THC & 0.20 & 0.018 & 0.014 & 0.012 \\ 
 \hline 
 & THA & 0.29 & 0.011 & 0.013 & 0.009 \\ 
\crowcolorC \cellcolor{white}\multirow{-2}{*}{$g_{R}^{\tau}$} & THC & 0.28 & 0.026 & 0.02 & 0.019 \\ 
 \hline 
 & THA & 0.73 & 0.019 & 0.014 & 0.012 \\ 
\crowcolorC \cellcolor{white}\multirow{-2}{*}{$g_{L}^{u=c}$} & THC & 0.73 & 0.045 & 0.05 & 0.048 \\ 
 \hline 
 & THA & 2.9 & 0.052 & 0.032 & 0.031 \\ 
\crowcolorC \cellcolor{white}\multirow{-2}{*}{$g_{R}^{u=c}$} & THC & 2.9 & 0.074 & 0.077 & 0.075 \\ 
 \hline 
 & THA & 0.57 & 0.018 & 0.011 & 0.01 \\ 
\crowcolorC \cellcolor{white}\multirow{-2}{*}{$g_{L}^{d=s}$} & THC & 0.57 & 0.037 & 0.04 & 0.039 \\ 
 \hline 
 & THA & 16 & 0.39 & 0.21 & 0.20 \\ 
\crowcolorC \cellcolor{white}\multirow{-2}{*}{$g_{R}^{d=s}$} & THC & 16 & 0.48 & 0.40 & 0.40 \\ 
 \hline 
 & THA & 0.37 & 0.0088 & 0.0094 & 0.0063 \\ 
\crowcolorC \cellcolor{white}\multirow{-2}{*}{$g_{L}^{b}$} & THC & 0.38 & 0.015 & 0.018 & 0.017 \\ 
 \hline 
 & THA & 11 & 0.20 & 0.15 & 0.15 \\ 
\crowcolorC \cellcolor{white}\multirow{-2}{*}{$g_{R}^{b}$} & THC & 12 & 0.25 & 0.38 & 0.38 \\ 
\arrayrulecolor{black}\bottomrule
\end{tabular}
\caption{Same as Table~\ref{tab:eft-global-EWgi1}, for additional combinations of collider results involving LHeC, LCF or LEP3.
\label{tab:eft-global-EWgi1Comb}
}
\end{table}

\begin{sidewaystable}[thp]
\begin{center}
\scriptsize
\begin{tabular}{ c | c | c | c | ccc | ccc | c | ccc | c}
\toprule
        &     & &\multicolumn{12}{c}{HL-LHC +}  \\
& TH & HL-LHC & LHeC & \multicolumn{3}{c|}{LCF} & \multicolumn{3}{c|}{CLIC} & LEP3 & \multicolumn{2}{c}{FCC--ee} & FCC & MuC$_{10}$ \\
& scenario    &  &  & 250 & 550 & 1000 & 380 & 1500 & 3000 &  & 240 & 365 &  &  \\ 
\midrule\midrule
 & THA & 1.2~(15) & 1.3~(15) & 3.7~(48) & 16~(53) & 27~(58) & 8.5~(91) & 28~(130) & 34~(200) & 4.3~(120) & 5.1~(160) & 10~(160) & 12~(160) & 130~(160) \\ 
\crowcolorC \cellcolor{white}\multirow{-2}{*}{$O_{\phi l}^{(1)}$} & THC & 1.2~(14) & 1.3~(15) & 3.7~(42) & 15~(47) & 22~(54) & 6.7~(90) & 24~(130) & 32~(190) & 3.9~(71) & 4.4~(73) & 9.3~(74) & 11~(74) & 130~(160) \\ 
 \hline 
 & THA & 0.96~(7.7) & 0.97~(7.8) & 0.99~(34) & 2.0~(34) & 2.1~(34) & 1.2~(10) & 1.4~(11) & 2.8~(14) & 1.2~(120) & 1.2~(160) & 1.7~(160) & 1.6~(160) & 1.4~(7.8) \\ 
\crowcolorC \cellcolor{white}\multirow{-2}{*}{$O_{\phi l,33}^{(1)}$} & THC & 0.96~(7.7) & 0.96~(7.7) & 0.99~(29) & 1.8~(29) & 1.9~(29) & 1.2~(10) & 1.4~(11) & 2.7~(13) & 1.1~(68) & 1.2~(68) & 1.6~(69) & 1.6~(69) & 1.4~(7.8) \\ 
 \hline 
 & THA & 1.6~(20) & 1.7~(22) & 4.8~(55) & 21~(58) & 31~(59) & 7.8~(94) & 23~(130) & 27~(200) & 4.8~(130) & 5.5~(170) & 10~(180) & 14~(170) & 93~(180) \\ 
\crowcolorC \cellcolor{white}\multirow{-2}{*}{$O_{\phi l}^{(3)}$} & THC & 1.7~(20) & 1.7~(22) & 4.2~(50) & 21~(53) & 29~(55) & 6.7~(93) & 22~(130) & 27~(200) & 4.5~(77) & 5.0~(78) & 9.3~(84) & 13~(84) & 92~(180) \\ 
 \hline 
 & THA & 1.2~(8.3) & 1.3~(8.3) & 2.9~(39) & 5.2~(39) & 5.7~(39) & 1.8~(14) & 3.7~(15) & 7.9~(16) & 2.9~(130) & 3.3~(170) & 5.5~(170) & 5.9~(170) & 2.0~(8.4) \\ 
\crowcolorC \cellcolor{white}\multirow{-2}{*}{$O_{\phi l,33}^{(1)}$} & THC & 1.2~(8.3) & 1.3~(8.3) & 2.6~(34) & 4.5~(34) & 5.0~(34) & 1.7~(14) & 3.7~(15) & 7.7~(16) & 2.8~(70) & 3.0~(71) & 4.4~(72) & 4.6~(71) & 2.0~(8.4) \\ 
 \hline 
 & THA & 1.2~(13) & 1.2~(13) & 2.8~(50) & 12~(56) & 23~(60) & 7.6~(86) & 20~(130) & 29~(200) & 3.2~(120) & 3.9~(160) & 9.5~(160) & 11~(160) & 85~(170) \\ 
\crowcolorC \cellcolor{white}\multirow{-2}{*}{$O_{\phi e}$} & THC & 1.2~(13) & 1.3~(13) & 2.3~(40) & 11~(46) & 19~(53) & 7.3~(85) & 19~(130) & 28~(200) & 2.7~(69) & 3.1~(71) & 7.8~(71) & 8.8~(71) & 84~(170) \\ 
 \hline 
 & THA & 0.96~(7.5) & 0.97~(7.5) & 0.97~(35) & 2.1~(35) & 3.3~(35) & 0.97~(10) & 2.8~(11) & 3.3~(13) & 1.2~(120) & 1.2~(150) & 1.8~(160) & 1.7~(150) & 1.1~(7.5) \\ 
\crowcolorC \cellcolor{white}\multirow{-2}{*}{$O_{\phi e,33}$} & THC & 0.96~(7.4) & 0.96~(7.5) & 0.97~(29) & 2.0~(29) & 3.2~(29) & 0.98~(10) & 2.8~(11) & 3.3~(13) & 1.2~(65) & 1.2~(66) & 1.7~(66) & 1.7~(66) & 1.1~(7.5) \\ 
 \hline 
 & THA & 2.0~(4.3) & 2.0~(4.4) & 2.5~(24) & 3.3~(24) & 4.4~(24) & 2.2~(10) & 2.5~(13) & 2.5~(18) & 2.6~(32) & 2.6~(35) & 3.0~(36) & 5.8~(36) & 2.6~(4.7) \\ 
\crowcolorC \cellcolor{white}\multirow{-2}{*}{$O_{\phi q}^{(1)}$} & THC & 2.0~(4.3) & 2.0~(4.4) & 2.5~(16) & 3.1~(16) & 3.9~(16) & 2.2~(9.9) & 2.4~(13) & 2.5~(17) & 2.5~(18) & 2.6~(19) & 2.9~(19) & 4.5~(20) & 2.6~(4.7) \\ 
 \hline 
 & THA & 0.74~(11) & 0.75~(12) & 0.86~(37) & 1.5~(38) & 2.8~(40) & 1.0~(24) & 2.1~(32) & 3.0~(47) & 1.1~(54) & 1.1~(58) & 1.6~(63) & 2.4~(63) & 5.0~(14) \\ 
\crowcolorC \cellcolor{white}\multirow{-2}{*}{$O_{\phi q,33}^{(1)}$} & THC & 0.74~(11) & 0.75~(12) & 0.84~(27) & 1.5~(30) & 2.7~(34) & 1.0~(21) & 2.1~(30) & 3.0~(46) & 1.1~(30) & 1.2~(31) & 1.5~(35) & 2.3~(34) & 4.9~(14) \\ 
 \hline 
 & THA & 4.2~(12) & 4.3~(13) & 5.5~(42) & 8.5~(42) & 9.1~(42) & 6.1~(22) & 7.1~(26) & 7.9~(33) & 6.1~(52) & 6.5~(55) & 9.6~(58) & 23~(59) & 7.0~(13) \\ 
\crowcolorC \cellcolor{white}\multirow{-2}{*}{$O_{\phi q}^{(3)}$} & THC & 4.2~(12) & 4.3~(13) & 5.5~(38) & 8.0~(38) & 8.5~(38) & 6.1~(21) & 7.0~(25) & 7.8~(33) & 6.0~(40) & 6.3~(40) & 8.4~(41) & 23~(45) & 7.0~(13) \\ 
 \hline 
 & THA & 0.93~(8.) & 0.94~(8.3) & 1.0~(36) & 3.0~(37) & 4.5~(37) & 1.3~(17) & 3.7~(20) & 4.4~(26) & 1.2~(55) & 1.3~(56) & 1.8~(57) & 2.8~(57) & 5.2~(12) \\ 
\crowcolorC \cellcolor{white}\multirow{-2}{*}{$O_{\phi q,33}^{(3)}$} & THC & 0.93~(8.) & 0.94~(8.3) & 1.0~(26) & 2.9~(26) & 4.4~(27) & 1.3~(16) & 3.7~(19) & 4.3~(25) & 1.2~(28) & 1.3~(28) & 1.7~(29) & 2.7~(29) & 5.2~(11) \\ 
 \hline 
 & THA & 0.9~(5.5) & 0.92~(5.7) & 1.3~(22) & 1.7~(22) & 2.5~(23) & 1.2~(14) & 1.9~(18) & 2.1~(25) & 1.5~(29) & 1.6~(39) & 1.9~(40) & 3.7~(40) & 1.5~(5.7) \\ 
\crowcolorC \cellcolor{white}\multirow{-2}{*}{$O_{\phi u}$} & THC & 0.89~(5.5) & 0.92~(5.6) & 1.2~(20) & 1.6~(20) & 2.5~(20) & 1.1~(14) & 1.9~(18) & 2.0~(24) & 1.4~(22) & 1.5~(23) & 1.8~(23) & 3.3~(23) & 1.5~(5.7) \\ 
 \hline 
 & THA & 0.88~(3.8) & 0.89~(3.9) & 1.1~(14) & 1.4~(15) & 1.8~(14) & 0.96~(9.3) & 1.2~(12) & 1.4~(17) & 1.1~(19) & 1.1~(23) & 1.3~(24) & 2.2~(24) & 1.2~(4.) \\ 
\crowcolorC \cellcolor{white}\multirow{-2}{*}{$O_{\phi d}$} & THC & 0.88~(3.8) & 0.9~(3.9) & 1.1~(13) & 1.3~(13) & 1.7~(13) & 0.96~(9.) & 1.2~(12) & 1.4~(17) & 1.1~(14) & 1.1~(15) & 1.3~(15) & 2.0~(15) & 1.2~(4.) \\ 
 \hline 
 & THA & 0.94~(3.1) & 0.94~(3.2) & 0.96~(17) & 1.6~(17) & 2.5~(17) & 0.96~(7.8) & 2.1~(9.4) & 2.4~(13) & 0.96~(24) & 0.97~(29) & 1.4~(29) & 1.4~(29) & 0.94~(3.2) \\ 
\crowcolorC \cellcolor{white}\multirow{-2}{*}{$O_{\phi d,33}$} & THC & 0.94~(3.1) & 0.94~(3.2) & 0.96~(14) & 1.6~(14) & 2.4~(14) & 0.96~(7.7) & 2.1~(9.3) & 2.4~(12) & 0.97~(13) & 0.97~(14) & 1.4~(14) & 1.4~(14) & 0.95~(3.2) \\ 
 \hline 
 & THA & 2.0~(31) & 2.1~(33) & 2.7~(53) & 17~(58) & 21~(64) & 11~(50) & 17~(62) & 20~(87) & 3.1~(89) & 3.5~(120) & 11~(140) & 13~(140) & 12~(40) \\ 
\crowcolorC \cellcolor{white}\multirow{-2}{*}{$O_{\phi WB}$} & THC & 2.0~(31) & 2.1~(32) & 2.7~(48) & 16~(53) & 19~(61) & 9.6~(42) & 15~(58) & 17~(85) & 2.6~(64) & 2.9~(67) & 8.7~(78) & 9.9~(80) & 12~(39) \\ 
 \hline 
 & THA & 1.1~(14) & 1.1~(16) & 1.4~(38) & 5.8~(40) & 9.6~(43) & 2.5~(34) & 9.5~(40) & 15~(52) & 1.8~(46) & 2.0~(59) & 3.0~(75) & 4.0~(75) & 4.8~(16) \\ 
\crowcolorC \cellcolor{white}\multirow{-2}{*}{$O_{\phi D}$} & THC & 1.1~(14) & 1.1~(16) & 1.4~(27) & 5.3~(31) & 8.7~(36) & 2.3~(29) & 9.2~(36) & 14~(51) & 1.8~(31) & 1.9~(33) & 3.0~(41) & 3.9~(41) & 4.7~(16) \\ 
\bottomrule
\end{tabular}
\caption{\label{tab:eft-global-OiEW1} 
68\% probability limits on the interaction scale associated to the different operators contributing to EWPOs presented in Figure 3.5 of Ref.~\cite{deBlas:2025PhysicsBriefingBook}. Compared to that reference, where only results in the {\it aggressive} (THA) theory scenario were presented, here we also report the limits in the {\it conservative} theory scenario (THC).
}
\end{center}
\end{sidewaystable}

\begin{table}[ht]
\centering
\setlength\tabcolsep{4.5pt}
\setlength\tabcolsep{2.5pt}
\begin{tabular}{ c | c | cccc }
\toprule
& TH & \multicolumn{4}{c}{HL-LHC+} \\
 & scenario & LHeC+FCC--hh & LCF$_{1000}$+MuC$_{10}$ & LEP3+FCC--hh & LEP3+MuC$_{10}$ \\ 
\midrule
\midrule
 & THA & 1.5~(15) & 130~(15) & 5.2~(48) & 130~(53) \\ 
\crowcolorC \cellcolor{white}\multirow{-2}{*}{$O_{\phi l}^{(1)}$} & THC & 1.5~(14) & 130~(15) & 4.8~(42) & 130~(47) \\ 
 \hline 
 & THA & 0.96~(7.7) & 3.1~(7.8) & 1.1~(34) & 2.2~(34) \\ 
\crowcolorC \cellcolor{white}\multirow{-2}{*}{$O_{\phi l,33}^{(1)}$} & THC & 0.97~(7.7) & 3.0~(7.7) & 1.1~(29) & 2.2~(29) \\ 
 \hline 
 & THA & 2.8~(20) & 94~(22) & 5.3~(55) & 93~(58) \\ 
\crowcolorC \cellcolor{white}\multirow{-2}{*}{$O_{\phi l}^{(3)}$} & THC & 2.8~(20) & 93~(22) & 5.1~(50) & 92~(53) \\ 
 \hline 
 & THA & 1.5~(8.3) & 10~(8.3) & 3.4~(39) & 5.2~(39) \\ 
\crowcolorC \cellcolor{white}\multirow{-2}{*}{$O_{\phi l,33}^{(1)}$} & THC & 1.5~(8.3) & 10~(8.3) & 3.1~(34) & 4.9~(34) \\ 
 \hline 
 & THA & 1.3~(13) & 86~(13) & 3.8~(50) & 85~(56) \\ 
\crowcolorC \cellcolor{white}\multirow{-2}{*}{$O_{\phi e}$} & THC & 1.3~(13) & 85~(13) & 3.4~(40) & 85~(46) \\ 
 \hline 
 & THA & 0.96~(7.5) & 4.6~(7.5) & 1.2~(35) & 1.9~(35) \\ 
\crowcolorC \cellcolor{white}\multirow{-2}{*}{$O_{\phi e,33}$} & THC & 0.95~(7.4) & 4.4~(7.5) & 1.2~(29) & 1.9~(29) \\ 
 \hline 
 & THA & 2.5~(4.3) & 5.7~(4.4) & 4.6~(24) & 4.1~(24) \\ 
\crowcolorC \cellcolor{white}\multirow{-2}{*}{$O_{\phi q}^{(1)}$} & THC & 2.5~(4.3) & 4.7~(4.4) & 4.0~(16) & 3.8~(16) \\ 
 \hline 
 & THA & 1.4~(11) & 7.0~(12) & 1.7~(37) & 5.7~(38) \\ 
\crowcolorC \cellcolor{white}\multirow{-2}{*}{$O_{\phi q,33}^{(1)}$} & THC & 1.4~(11) & 6.8~(12) & 1.6~(27) & 5.6~(30) \\ 
 \hline 
 & THA & 18~(12) & 15~(13) & 20~(42) & 10~(42) \\ 
\crowcolorC \cellcolor{white}\multirow{-2}{*}{$O_{\phi q}^{(3)}$} & THC & 18~(12) & 15~(13) & 20~(38) & 9.1~(38) \\ 
 \hline 
 & THA & 2.1~(8.) & 10~(8.3) & 2.3~(36) & 6.0~(37) \\ 
\crowcolorC \cellcolor{white}\multirow{-2}{*}{$O_{\phi q,33}^{(3)}$} & THC & 2.1~(8.) & 9.6~(8.3) & 2.3~(26) & 5.8~(26) \\ 
 \hline 
 & THA & 1.9~(5.5) & 4.0~(5.7) & 3.1~(22) & 2.3~(22) \\ 
\crowcolorC \cellcolor{white}\multirow{-2}{*}{$O_{\phi u}$} & THC & 1.9~(5.5) & 3.7~(5.6) & 3.0~(20) & 2.1~(20) \\ 
 \hline 
 & THA & 1.3~(3.8) & 2.5~(3.9) & 1.8~(14) & 1.7~(15) \\ 
\crowcolorC \cellcolor{white}\multirow{-2}{*}{$O_{\phi d}$} & THC & 1.3~(3.8) & 2.4~(3.9) & 1.7~(13) & 1.6~(13) \\ 
 \hline 
 & THA & 0.95~(3.1) & 3.6~(3.2) & 1.0~(17) & 1.7~(17) \\ 
\crowcolorC \cellcolor{white}\multirow{-2}{*}{$O_{\phi d,33}$} & THC & 0.95~(3.1) & 3.5~(3.2) & 1.0~(14) & 1.7~(14) \\ 
 \hline 
 & THA & 2.9~(31) & 33~(33) & 4.0~(53) & 23~(58) \\ 
\crowcolorC \cellcolor{white}\multirow{-2}{*}{$O_{\phi WB}$} & THC & 2.9~(31) & 31~(32) & 3.4~(48) & 23~(53) \\ 
 \hline 
 & THA & 2.0~(14) & 13~(16) & 2.5~(38) & 5.7~(40) \\ 
\crowcolorC \cellcolor{white}\multirow{-2}{*}{$O_{\phi D}$} & THC & 2.0~(14) & 13~(16) & 2.5~(27) & 5.4~(31) \\ 
\arrayrulecolor{black}\bottomrule
\end{tabular}
\caption{Same as Table~\ref{tab:eft-global-OiEW1}, for additional combinations of collider results involving LHeC, LCF or LEP3.
\label{tab:eft-global-OiEW1Comb}
}
\end{table}

\begin{sidewaystable}[thp]
\begin{center}
\scriptsize
\begin{tabular}{ c | c | c | c | ccc | ccc | c | ccc | c}
\toprule
        &  &   &\multicolumn{12}{c}{HL-LHC +}  \\
& TH & HL-LHC & LHeC & \multicolumn{3}{c|}{LCF} & \multicolumn{3}{c|}{CLIC} & LEP3 & \multicolumn{2}{c}{FCC--ee} & FCC & MuC$_{10}$ \\
& scenario    &  &  & 250 & 550 & 1000 & 380 & 1500 & 3000 &  & 240 & 365 &  &  \\  
\midrule\midrule
 & THA & 0.63~(5.4) & 0.65~(5.5) & 1.1~(39) & 7.6~(71) & 16~(99) & 3.1~(43) & 14~(93) & 41~(140) & 1.1~(51) & 1.2~(66) & 4.0~(69) & 6.3~(69) & 130~(300) \\ 
\crowcolorC \cellcolor{white}\multirow{-2}{*}{$O_{Ql}^{(-)}$} & THC & 0.63~(5.4) & 0.65~(5.4) & 0.98~(31) & 7.5~(65) & 16~(95) & 3.0~(43) & 14~(92) & 41~(140) & 1.1~(32) & 1.2~(38) & 4.0~(46) & 6.0~(46) & 130~(310) \\ 
 \hline 
 & THA & 0.61~(7.7) & 0.62~(8.3) & 1.5~(55) & 9.2~(95) & 18~(130) & 3.6~(53) & 17~(120) & 49~(180) & 1.6~(59) & 1.7~(75) & 3.7~(83) & 5.4~(83) & 350~(440) \\ 
\crowcolorC \cellcolor{white}\multirow{-2}{*}{$O_{Ql}^{(3)}$} & THC & 0.61~(7.7) & 0.62~(8.2) & 1.4~(43) & 8.9~(85) & 18~(130) & 3.5~(51) & 16~(120) & 49~(180) & 1.6~(40) & 1.7~(51) & 3.4~(62) & 5.0~(62) & 350~(440) \\ 
 \hline 
 & THA & 0.53~(4.6) & 0.53~(4.6) & 0.89~(17) & 4.2~(46) & 15~(61) & 2.9~(32) & 12~(54) & 29~(80) & 1.1~(42) & 1.2~(55) & 4.3~(56) & 5.1~(56) & 140~(280) \\ 
\crowcolorC \cellcolor{white}\multirow{-2}{*}{$O_{te}$} & THC & 0.53~(4.6) & 0.53~(4.6) & 0.78~(14) & 4.1~(45) & 15~(62) & 2.9~(32) & 12~(54) & 29~(80) & 0.95~(24) & 1.0~(24) & 4.1~(26) & 4.9~(26) & 140~(280) \\ 
 \hline 
 & THA & 0.49~(4.9) & 0.55~(5.) & 1.4~(16) & 6.1~(43) & 20~(57) & 4.0~(33) & 17~(65) & 41~(97) & 1.3~(42) & 1.4~(56) & 5.5~(56) & 6.2~(56) & 220~(250) \\ 
\crowcolorC \cellcolor{white}\multirow{-2}{*}{$O_{tl}$} & THC & 0.49~(4.9) & 0.56~(5.) & 1.3~(14) & 5.9~(43) & 20~(57) & 3.9~(33) & 17~(65) & 41~(97) & 1.3~(24) & 1.4~(25) & 5.3~(28) & 6.1~(28) & 220~(250) \\ 
 \hline 
 & THA & 0.52~(4.8) & 0.52~(4.8) & 2.9~(34) & 11~(63) & 19~(85) & 3.1~(38) & 18~(59) & 36~(88) & 2.8~(45) & 2.9~(58) & 3.7~(59) & 5.0~(59) & 190~(270) \\ 
\crowcolorC \cellcolor{white}\multirow{-2}{*}{$O_{Q e}$} & THC & 0.53~(4.8) & 0.53~(4.8) & 2.9~(32) & 11~(62) & 18~(85) & 3.1~(38) & 18~(59) & 36~(88) & 2.8~(26) & 3.0~(29) & 3.6~(31) & 4.6~(31) & 190~(270) \\ 
 \hline 
 & THA & 0.58~(11) & 0.59~(12) & 0.67~(37) & 1.4~(38) & 2.5~(40) & 0.83~(24) & 2.0~(32) & 2.7~(47) & 0.81~(54) & 0.84~(58) & 1.2~(63) & 1.9~(63) & 6.1~(14) \\ 
\crowcolorC \cellcolor{white}\multirow{-2}{*}{$O_{\phi Q}^{(-)}$} & THC & 0.58~(11) & 0.59~(12) & 0.65~(27) & 1.4~(30) & 2.5~(34) & 0.8~(21) & 1.9~(30) & 2.7~(46) & 0.81~(30) & 0.89~(31) & 1.1~(34) & 1.8~(34) & 6.1~(14) \\ 
 \hline 
 & THA & 0.77~(9.7) & 0.78~(11) & 0.87~(25) & 1.5~(28) & 2.6~(33) & 1.2~(25) & 2.0~(32) & 2.9~(45) & 0.91~(31) & 0.95~(39) & 1.5~(50) & 1.7~(50) & 4.1~(11) \\ 
\crowcolorC \cellcolor{white}\multirow{-2}{*}{$O_{\phi t}$} & THC & 0.76~(9.6) & 0.78~(11) & 0.86~(19) & 1.4~(24) & 2.5~(30) & 1.2~(22) & 2.0~(30) & 2.9~(45) & 0.92~(21) & 1.0~(22) & 1.5~(28) & 1.7~(28) & 4.1~(11) \\ 
 \hline 
 & THA & 0.93~(13) & 0.94~(14) & 1.0~(51) & 3.0~(52) & 4.5~(53) & 1.3~(29) & 3.7~(36) & 4.4~(51) & 1.2~(77) & 1.3~(80) & 1.8~(83) & 2.8~(83) & 5.2~(15) \\ 
\crowcolorC \cellcolor{white}\multirow{-2}{*}{$O_{\phi Q}^{(3)}$} & THC & 0.93~(13) & 0.94~(14) & 1.0~(38) & 2.9~(39) & 4.4~(42) & 1.3~(26) & 3.7~(34) & 4.3~(50) & 1.2~(40) & 1.3~(41) & 1.7~(44) & 2.7~(44) & 5.2~(15) \\ 
 \hline 
 & THA & 0.64~(1.9) & 0.64~(2.0) & 0.71~(2.8) & 1.2~(3.1) & 1.4~(3.4) & 0.78~(2.9) & 1.0~(3.1) & 1.3~(3.3) & 0.71~(2.6) & 0.77~(3.2) & 0.85~(3.3) & 2.0~(4.9) & 1.2~(3.3) \\ 
\crowcolorC \cellcolor{white}\multirow{-2}{*}{$O_{t \phi}$} & THC & 0.63~(1.9) & 0.65~(2.0) & 0.7~(2.6) & 1.2~(2.9) & 1.4~(3.1) & 0.78~(2.8) & 1.0~(2.9) & 1.3~(3.0) & 0.7~(2.5) & 0.76~(2.9) & 0.85~(3.0) & 2.0~(4.7) & 1.1~(3.0) \\ 
 \hline 
 & THA & 1.1~(19) & 1.1~(20) & 1.2~(25) & 7.5~(38) & 8.6~(41) & 5.1~(33) & 8.2~(35) & 12~(41) & 1.2~(38) & 1.3~(49) & 4.0~(57) & 4.6~(60) & 6.3~(27) \\ 
\crowcolorC \cellcolor{white}\multirow{-2}{*}{$O_{tZ}$} & THC & 1.1~(19) & 1.1~(20) & 1.2~(23) & 7.4~(38) & 8.7~(41) & 5.1~(32) & 8.2~(35) & 12~(41) & 1.2~(28) & 1.2~(30) & 4.0~(36) & 4.5~(40) & 6.2~(26) \\ 
 \hline 
 & THA & 0.95~(21) & 0.99~(21) & 1.3~(28) & 5.8~(42) & 6.6~(45) & 5.2~(35) & 7.4~(39) & 12~(47) & 1.3~(43) & 1.6~(57) & 4.7~(67) & 5.5~(69) & 5.6~(29) \\ 
\crowcolorC \cellcolor{white}\multirow{-2}{*}{$O_{tW}$} & THC & 0.94~(21) & 1.0~(21) & 1.2~(26) & 5.7~(41) & 6.7~(44) & 5.1~(34) & 7.4~(38) & 11~(46) & 1.3~(32) & 1.5~(34) & 4.7~(41) & 5.5~(45) & 5.6~(28) \\ 
\bottomrule
\end{tabular}
\caption{\label{tab:eft-global-OiTop1} 
68\% probability limits on the interaction scale associated to $\ell^+ \ell^- t\bar{t}$ interactions and operators modifying the EW and Higgs couplings of the top-quark, as shown in Figure 3.8 of Ref.~\cite{deBlas:2025PhysicsBriefingBook}. Compared to that reference, where only results in the {\it aggressive} (THA) theory scenario were presented, here we also report the limits in the {\it conservative} theory scenario (THC).
}
\end{center}
\end{sidewaystable}

\begin{table}[ht]
\centering
\setlength\tabcolsep{4.5pt}
\setlength\tabcolsep{2.5pt}
\begin{tabular}{ c | c | cccc }
\toprule
& TH & \multicolumn{4}{c}{HL-LHC+} \\
 & scenario & LHeC+FCC--hh & LCF$_{1000}$+MuC$_{10}$ & LEP3+FCC--hh & LEP3+MuC$_{10}$ \\ 
\midrule
\midrule
 & THA & 0.7~(5.4) & 130~(5.5) & 1.5~(39) & 130~(71) \\ 
\crowcolorC \cellcolor{white}\multirow{-2}{*}{$O_{Ql}^{(-)}$} & THC & 0.7~(5.4) & 130~(5.4) & 1.4~(31) & 130~(65) \\ 
 \hline 
 & THA & 1.0~(7.7) & 350~(8.3) & 2.0~(55) & 340~(95) \\ 
\crowcolorC \cellcolor{white}\multirow{-2}{*}{$O_{Ql}^{(3)}$} & THC & 1.0~(7.7) & 350~(8.2) & 1.9~(43) & 350~(85) \\ 
 \hline 
 & THA & 0.56~(4.6) & 140~(4.6) & 1.3~(17) & 140~(46) \\ 
\crowcolorC \cellcolor{white}\multirow{-2}{*}{$O_{te}$} & THC & 0.56~(4.6) & 150~(4.6) & 1.2~(14) & 140~(45) \\ 
 \hline 
 & THA & 0.58~(4.9) & 220~(5.0) & 1.6~(16) & 220~(43) \\ 
\crowcolorC \cellcolor{white}\multirow{-2}{*}{$O_{tl}$} & THC & 0.57~(4.9) & 210~(5.0) & 1.5~(14) & 220~(43) \\ 
 \hline 
 & THA & 0.6~(4.8) & 190~(4.8) & 3.4~(34) & 190~(63) \\ 
\crowcolorC \cellcolor{white}\multirow{-2}{*}{$O_{Q e}$} & THC & 0.6~(4.8) & 190~(4.8) & 3.3~(32) & 190~(62) \\ 
 \hline 
 & THA & 1.2~(11) & 7.9~(12) & 1.4~(37) & 6.8~(38) \\ 
\crowcolorC \cellcolor{white}\multirow{-2}{*}{$O_{\phi Q}^{(-)}$} & THC & 1.2~(11) & 7.8~(12) & 1.4~(27) & 6.8~(30) \\ 
 \hline 
 & THA & 1.1~(9.7) & 5.4~(11) & 1.2~(25) & 4.7~(28) \\ 
\crowcolorC \cellcolor{white}\multirow{-2}{*}{$O_{\phi t}$} & THC & 1.1~(9.6) & 5.3~(11) & 1.2~(19) & 4.6~(24) \\ 
 \hline 
 & THA & 2.1~(13) & 10~(14) & 2.3~(51) & 6.0~(52) \\ 
\crowcolorC \cellcolor{white}\multirow{-2}{*}{$O_{\phi Q}^{(3)}$} & THC & 2.1~(13) & 9.6~(14) & 2.3~(38) & 5.8~(39) \\ 
 \hline 
 & THA & 1.7~(1.9) & 1.7~(2.0) & 1.8~(2.8) & 1.2~(3.1) \\ 
\crowcolorC \cellcolor{white}\multirow{-2}{*}{$O_{t \phi}$} & THC & 1.7~(1.9) & 1.6~(2.0) & 1.8~(2.6) & 1.2~(2.9) \\ 
 \hline 
 & THA & 1.1~(19) & 12~(20) & 1.3~(25) & 6.4~(38) \\ 
\crowcolorC \cellcolor{white}\multirow{-2}{*}{$O_{tZ}$} & THC & 1.1~(19) & 12~(20) & 1.3~(23) & 6.3~(38) \\ 
 \hline 
 & THA & 1.4~(21) & 8.5~(21) & 1.6~(28) & 5.7~(42) \\ 
\crowcolorC \cellcolor{white}\multirow{-2}{*}{$O_{tW}$} & THC & 1.4~(21) & 8.5~(21) & 1.6~(26) & 5.7~(41) \\ 
\arrayrulecolor{black}\bottomrule
\end{tabular}
\caption{Same as Table~\ref{tab:eft-global-OiTop1}, for additional combinations of collider results involving LHeC, LCF or LEP3.
\label{tab:eft-global-OiTop1Comb}
}
\end{table}

\clearpage

\begin{sidewaystable}[thp]
\begin{center}
\scriptsize
\begin{tabular}{ c | c | c | c | ccc | ccc | c | ccc | c}
\toprule
        &  &   &\multicolumn{12}{c}{HL-LHC +}  \\
& TH & HL-LHC & LHeC & \multicolumn{3}{c|}{LCF} & \multicolumn{3}{c|}{CLIC} & LEP3 & \multicolumn{2}{c}{FCC--ee} & FCC & MuC$_{10}$ \\
& scenario    &  &  & 250 & 550 & 1000 & 380 & 1500 & 3000 &  & 240 & 365 &  &  \\  
\midrule\midrule
 & THA & 3.0~(13) & 4.4~(14) & 9.2~(28) & 12~(31) & 12~(32) & 13~(60) & 19~(90) & 21~(140) & 8.2~(50) & 9.9~(66) & 12~(75) & 17~(75) & 6.6~(100) \\ 
\crowcolorC \cellcolor{white}\multirow{-2}{*}{$O_{B}$} & THC & 3.0~(13) & 4.5~(14) & 9.1~(26) & 12~(27) & 12~(30) & 13~(58) & 19~(91) & 20~(140) & 8.0~(35) & 9.8~(37) & 11~(43) & 16~(43) & 6.5~(100) \\ 
 \hline 
 & THA & 2.2~(11) & 3.4~(11) & 6.9~(22) & 9.1~(25) & 9.5~(26) & 9.9~(51) & 15~(72) & 17~(110) & 6.1~(36) & 7.4~(49) & 8.7~(55) & 13~(56) & 5.8~(100) \\ 
\crowcolorC \cellcolor{white}\multirow{-2}{*}{$O_{W}$} & THC & 2.2~(11) & 3.4~(11) & 6.8~(20) & 8.9~(23) & 9.3~(25) & 9.9~(50) & 15~(72) & 17~(110) & 5.9~(26) & 7.3~(27) & 8.6~(32) & 12~(33) & 5.7~(100) \\ 
 \hline 
 & THA & 13~(14) & 13~(14) & 26~(29) & 47~(51) & 70~(74) & 32~(43) & 57~(72) & 88~(110) & 23~(32) & 31~(43) & 36~(47) & 86~(89) & 230~(250) \\ 
\crowcolorC \cellcolor{white}\multirow{-2}{*}{$O_{2B}$} & THC & 13~(14) & 13~(14) & 19~(20) & 41~(41) & 66~(67) & 23~(39) & 56~(70) & 88~(110) & 16~(22) & 19~(24) & 25~(30) & 85~(88) & 230~(250) \\ 
 \hline 
 & THA & 23~(24) & 23~(24) & 27~(27) & 37~(38) & 53~(55) & 28~(43) & 54~(71) & 84~(110) & 31~(36) & 37~(42) & 38~(45) & 180~(180) & 230~(260) \\ 
\crowcolorC \cellcolor{white}\multirow{-2}{*}{$O_{2W}$} & THC & 23~(24) & 23~(24) & 26~(26) & 35~(36) & 52~(53) & 27~(42) & 54~(70) & 83~(110) & 24~(27) & 26~(28) & 27~(33) & 180~(180) & 240~(260) \\ 
 \hline 
 & THA & 4.9~(5.4) & 5.0~(5.4) & 6.4~(7.4) & 6.8~(8.2) & 7.2~(9.5) & 5.4~(6.6) & 6.8~(8.9) & 8.5~(13) & 8.2~(9.8) & 9.5~(12) & 10.0~(14) & 79~(80) & 5.1~(5.5) \\ 
\crowcolorC \cellcolor{white}\multirow{-2}{*}{$O_{2G}$} & THC & 4.9~(5.4) & 5.0~(5.4) & 5.5~(6.2) & 5.8~(7.3) & 6.4~(8.9) & 5.3~(6.4) & 6.8~(8.8) & 8.3~(13) & 5.9~(6.7) & 6.4~(7.3) & 6.8~(8.7) & 79~(79) & 5.1~(5.5) \\ 
\bottomrule
\end{tabular}
\caption{\label{tab:eft-global-OiSILH} 
68\% probability limits on universal two-fermion/two-boson and four-fermion contact interactions, as shown in Figure 3.9 of Ref.~\cite{deBlas:2025PhysicsBriefingBook}. Compared to that reference, where only results in the {\it aggressive} (THA) theory scenario were presented, here we also report the limits in the {\it conservative} theory scenario (THC).
}
\end{center}
\end{sidewaystable}

\begin{table}[ht]
\centering
\setlength\tabcolsep{4.5pt}
\setlength\tabcolsep{2.5pt}
\begin{tabular}{ c | c | cccc }
\toprule
& TH & \multicolumn{4}{c}{HL-LHC+} \\
 & scenario & LHeC+FCC--hh & LCF$_{1000}$+MuC$_{10}$ & LEP3+FCC--hh & LEP3+MuC$_{10}$ \\ 
\midrule
\midrule
 & THA & 6.6~(14) & 14~(100) & 12~(50) & 12~(100) \\ 
\crowcolorC \cellcolor{white}\multirow{-2}{*}{$O_{B}$} & THC & 6.6~(14) & 13~(100) & 11~(35) & 12~(100) \\ 
 \hline 
 & THA & 5.7~(19) & 11~(100) & 8.4~(37) & 9.0~(100) \\ 
\crowcolorC \cellcolor{white}\multirow{-2}{*}{$O_{W}$} & THC & 5.6~(19) & 10~(100) & 8.1~(27) & 8.8~(100) \\ 
 \hline 
 & THA & 84~(87) & 230~(250) & 85~(88) & 230~(250) \\ 
\crowcolorC \cellcolor{white}\multirow{-2}{*}{$O_{2B}$} & THC & 84~(88) & 230~(250) & 84~(87) & 230~(250) \\ 
 \hline 
 & THA & 180~(180) & 240~(260) & 180~(180) & 240~(260) \\ 
\crowcolorC \cellcolor{white}\multirow{-2}{*}{$O_{2W}$} & THC & 180~(180) & 230~(250) & 180~(180) & 230~(250) \\ 
 \hline 
 & THA & 79~(79) & 7.9~(9.5) & 79~(79) & 9.4~(9.8) \\ 
\crowcolorC \cellcolor{white}\multirow{-2}{*}{$O_{2G}$} & THC & 79~(80) & 7.2~(8.9) & 79~(79) & 6.0~(6.7) \\ 
\arrayrulecolor{black}\bottomrule
\end{tabular}
\caption{Same as Table~\ref{tab:eft-global-OiSILH}, for additional combinations of collider results involving LHeC, LCF or LEP3.
\label{tab:eft-global-OiSILHComb}
}
\end{table}

\clearpage

% -------------------------------------

\bibliographystyle{JHEP}{}
\bibliography{refs}

\end{document}